\documentclass[a4paper,11pt]{article}
\pdfoutput=1 

\usepackage{jheppub2} 
\usepackage{bbm,bm,graphicx,mathtools,color,slashed,hyperref}
\usepackage{amssymb,amsmath}
\usepackage{braket}
\usepackage{subfig}

\newcommand{\diff}{\mathrm{d}}

\newcommand{\tr}{\mathrm{tr}}
\newcommand{\im}{\mathrm{i}}

\newcommand{\rme}{\mathrm{e}}

\newcommand{\ee}{\,\mathrm{e}}

\newcommand{\ZZ}{\mathbb{Z}}
\newcommand{\RR}{\mathbb{R}}

\usepackage{tikz}
\usetikzlibrary{arrows.meta,decorations.markings,decorations.pathreplacing}
 
\usepackage{amsthm}

\theoremstyle{definition}
\newtheorem{claim}{Claim}

\newtheorem{proposition}{Proposition}

\newtheorem{definition}{Definition}

\usepackage{booktabs}   
\usepackage{array}      

\newcommand{\Td}{\hat T^2}
\newcommand{\Vh}{\hat V}
\newcommand{\Ah}{\hat A}
\newcommand{\Fh}{\hat F}

\newcommand{\lamh}{\hat\lambda}
\newcommand{\sbar}{\bar\sigma}
\newcommand{\etab}{\bar\eta}

\title{
Doubly-periodic instanton via the ADHM method and the self-dual monopole-center vortex chain
}

\author[1]{Yui Hayashi}
\emailAdd{yui.hayashi@yukawa.kyoto-u.ac.jp}
\affiliation[1]{Department of Physics, The University of Tokyo, 7-3-1 Hongo, Bunkyo-ku, Tokyo 113-0033, Japan}

\author[2]{and Yuya Tanizaki}
\emailAdd{yuya.tanizaki@yukawa.kyoto-u.ac.jp}
\affiliation[2]{Yukawa Institute for Theoretical Physics, Kyoto University, Kyoto 606-8502, Japan}

\preprint{YITP-26-144}

\abstract{
Calorons, i.e., instantons on $\mathbb{R}^3\times S^1$, with nontrivial asymptotic holonomy decompose into $N$ fundamental monopole constituents. This suggests that a doubly-periodic Yang--Mills instanton on $\mathbb{R}^2\times T^2$ may admit a description in terms of a periodic array of monopole constituents. Starting from the Atiyah--Drinfeld--Hitchin--Manin (ADHM) construction, we derive the $2$d Abelian Hitchin system with impurities as a dual description of the doubly-periodic instanton. 
We then give a prescription for reconstructing the instanton from the Green's function of this system, assuming a scaling ansatz for the ultraviolet divergences induced by the delta-function impurities. In the far-field limit with strongly anisotropic torus periods, we extend the analysis of Bruckmann and van Baal for multi-calorons and identify an alternating chain of BPS and Kaluza--Klein (KK) monopole constituents: 
The monopole-center vortex chain arises from the non-Abelian self-dual field especially when an 't Hooft twist cuts the torus into the $1/N$ fragment.
}

\begin{document}

\maketitle

\section{Introduction}
\label{sec:introduction}

Four-dimensional $SU(N)$ Yang–Mills theory (and QCD) on a small $\mathbb{R}^2\times T^2$ with 't Hooft twist exhibits color confinement in a weakly coupled semiclassical regime~\cite{Tanizaki:2022ngt, Tanizaki:2022plm, Hayashi:2023wwi, Hayashi:2024gxv, Hayashi:2024qkm, Bergner:2025qsm} (see also the preceding relevant works~\cite{RTN:1993ilw, Gonzalez-Arroyo:1995ynx, Yamazaki:2017ulc, Cox:2021vsa, Gonzalez-Arroyo:2023kqv} for the similar story on $\mathbb{R}\times T^3$). 
To this end, the center-vortex fractional instanton plays a pivotal role. 
However, explicit constructions of the full non-Abelian center-vortex fractional instanton have largely relied on numerical minimization of the lattice Wilson action~\cite{Gonzalez-Arroyo:1998hjb, Montero:1999by, Montero:2000pb} (see also \cite{GarciaPerez:1989gt, GarciaPerez:1992fj, Itou:2018wkm, DasilvaGolan:2022jlm, Wandler:2024hsq}), and the analytic structure is not yet fully understood except for the Abelian-like solution for the twisted $T^4$ with the tuned aspect ratios~\cite{tHooft:1981nnx}.

In previous works by the present authors~\cite{Hayashi:2024yjc, Hayashi:2024psa, Hayashi:2025doq}, we obtained hints toward an analytic understanding by considering the hierarchy of periods for $T^2_{\mathrm{twisted}}$, $\ell_3 \gg N\ell_4$. 
The 't Hooft twist leads to nontrivial asymptotic holonomies with the required noncommuting structure, so that the off-diagonal gauge fields become massive at distances larger than $N\ell_4$. 
In such a situation, the Kraan-van Baal-Lee-Lu-Yi (KvBLLY) caloron~\cite{Lee:1997vp, Lee:1998bb, Lee:1998vu, Kraan:1998kp, Kraan:1998pm, Kraan:1998sn} gives the $N$ types of the fundamental monopole constituents as fractional instantons.  
In \cite{Hayashi:2024yjc, Hayashi:2024psa, Hayashi:2025doq}, we have found that the alternating infinite chain of BPS and KK monopole constituents separated by $\ell_3$ describes the center-vortex fractional instanton by extraction of its minimal segment. This solves the self-dual Yang-Mills equation on $\mathbb{R}^2\times T^2$ with the 't~Hooft twisted boundary condition at least for the Abelianized regime. 

An important open question is whether the monopole-center vortex chain actually comes out of the fully non-Abelian self-dual solution (see Ref.~\cite{Dobozy:2026edc} for its numerical support), and it naturally motivates us to look back on how the KvBLLY caloron on $\mathbb{R}^3\times S^1$ has been constructed. 
The KvBLLY caloron is discovered by adopting the Atiyah-Drinfeld-Hitchin-Manin (ADHM) construction~\cite{Atiyah:1978ri} for the periodic instanton.
The analysis by K.-M.~Lee, C.-h.~Lu, and P.~Yi~\cite{Lee:1997vp, Lee:1998bb, Lee:1998vu} relies on its D-brane interpretation: Starting from $N$ D$4$-branes with one compact direction and $1$ D$0$-brane, they performed the T-duality on this setup, which gives $N$ D$3$-branes separated along the compact direction with $N$ segments of the D$1$-string connecting D$3$-branes.
Then, they wrote down and analyzed the corresponding Nahm equation~\cite{Nahm:1982jb} but with the impurity to derive the caloron solution with nontrivial asymptotic holonomies. 

The papers by Kraan and van Baal~\cite{Kraan:1998kp, Kraan:1998pm, Kraan:1998sn} instead use the ADHM construction more directly for constructing the periodic instanton: They started from the ADHM data for the charge-$k$ instanton on $\mathbb{R}^4$, took the formal $k\to \infty$ limit, 
where the infinite ADHM index is reinterpreted as the Fourier momentum along the compact direction. 
Then, the ADHM self-duality equation becomes the Nahm equation with the impurity after the Fourier transformation, and gives the self-dual caloron solutions.  
Later, this technique is extended to the construction of the multi-caloron solution by Bruckmann and van Baal~\cite{Bruckmann:2002vy, Bruckmann:2003ag}.

In this work, we extend the Bruckmann-Kraan-van Baal method to construct the doubly-periodic instanton starting from the ADHM data. 
When taking the formal $k\to \infty$ limit of the ADHM data for the charge-$k$ instanton on $\mathbb{R}^4$, we require that the ``$\infty$'' label has the $2$d $\mathbb{Z}^2$ lattice structure as the Fourier dual of $T^2$. 
As a consequence, the ADHM self-duality equation becomes the Hitchin equation with impurities, which has been derived by Kapustin and Sethi~\cite{Kapustin:1998pb} (see also \cite{Hamanaka:2001dr} for localized doubly-periodic instantons in noncommutative gauge theory). 
We formulate a prescription for reconstructing the doubly-periodic instanton from the Green's function of the Hitchin system, extending and reinterpreting the works of Jardim~\cite{Jardim:1999dx, Jardim:1999mi} and of Ford and Pawlowski~\cite{Ford:2002pa, Ford:2003vi, Ford:2005sq} (see also~\cite{Ford:2000zt}). This prescription relies on a scaling ansatz for the ultraviolet divergences induced by the $2$d delta-function impurities.
We then show that, in the far-field limit with strongly anisotropic torus periods, the analysis developed by Bruckmann and van Baal for multi-calorons~\cite{Bruckmann:2002vy} can be adapted to the impurity Hitchin system. 
The explicit far-field calculation provides a nontrivial check of the scaling ansatz and identifies the alternating chain of BPS and KK monopole constituents found in our previous works~\cite{Hayashi:2024yjc, Hayashi:2024psa, Hayashi:2025doq}.

\section{Brief review for the fractional instanton on \texorpdfstring{$\mathbb{R}^2\times (T^2)_{p/N\text{-twisted}}$}{R2xT2}}
\label{sec:review_centervortex_Abelian}

In this section, we give a brief review on the self-dual fractional instanton on $\mathbb{R}^2\times T^2$ with the 't~Hooft twisted boundary condition in the Abelianized regime based on Refs.~\cite{Hayashi:2024yjc, Hayashi:2024psa, Hayashi:2025doq}, and we shall see how the fractional instanton is realized as the self-dual monopole-center vortex chain. 

\subsection{Setup of the problem}

We consider the self-dual Yang-Mills equation on $\mathbb{R}^2\times T^2$ with the $p/N$ 't~Hooft twisted boundary condition with $\gcd(N,p)=1$. 
We write the coordinate as $x=(\bm{x},x_3,x_4)\in \mathbb{R}^2\times T^2$ with the identification 
\begin{equation}
    x_3\sim x_3+\ell_3, \quad x_4\sim x_4+\ell_4. 
\end{equation}
Let us introduce the $SU(N)$ clock and shift matrices $C,S$, that satisfy $SC=\omega CS$;
\begin{align}
    (C)_{nm}=\rme^{-\frac{\pi \im (N-1)}{N}}\rme^{\frac{2\pi\im}{N}n}\delta_{nm}, \quad 
    (S)_{nm}=\rme^{-\frac{\pi \im (N-1)}{N}}\delta_{n+1,m}. 
\end{align}
Here, the $SU(N)$ indices $n,m$ are understood in mod $N$. The overall $U(1)$ phase $\rme^{-\frac{\pi \im (N-1)}{N}}$ is introduced to make $C, S\in SU(N)$ for both even and odd $N$ with the same expression, and it plays no essential role in the following discussion. 

Throughout this paper, the gauge field $A=A_\mu\diff x^\mu$ is hermitian with the covariant derivative $D=\diff+\im A$. 
The field strength is $F=-\im D\wedge D= \diff A+\im A\wedge A$, and the Wilson loop is $\mathcal{P}\exp(-\im\int A)$.
The classical minimum of the Yang-Mills action on $\mathbb{R}^2\times T^2$ with the 't~Hooft twist $p/N$ is unique up to the gauge transformation, and the Polyakov loops $P_a=\mathcal{P}\exp(-\im\int_{x}^{x+\ell_a\hat{e}_a} A)$ for the classical minimum can be set as 
\begin{align}
    P_{3}=S^{-p}, \, P_{4}=C.
\end{align}
With $SC=\omega CS$ this gives $P_3P_4=\omega^{-p}P_4P_3$.
Any $SU(N)$ flat connection with the $p/N$ 't~Hooft twist is gauge equivalent to this configuration, and thus we do not lose any generality here (For the derivation, see, e.g., Ref.~\cite[Appendix~A]{Tanizaki:2022ngt}).
It is then convenient to take the algebraic gauge, where the gauge field (and also any adjoint fields) obeys
\begin{equation}
    A_\mu(x+\ell_a \hat{e}_a)= P_a\, A_\mu(x)\, P_a^{-1}\quad (a=3,4),
    \label{eq:algebraic_gauge}
\end{equation}
where $\hat{e}_a$ is the unit vector for the $a$-th direction. In this choice of the gauge, $A_{\mu=3,4}\to 0$ as $|\bm{x}|\to \infty$. 
Perturbative spectrum with this twisted boundary condition has the Kaluza-Klein mass gap, 
\begin{align}
    m^2_{(Nk_3+r_3, Nk_4+r_4)}=\left(\frac{2\pi}{N}\right)^2 \left\{\frac{(Nk_3+r_3)^2}{\ell_3^2}+\frac{(Nk_4+r_4)^2}{\ell_4^2}\right\}
    \label{eq:colormomentum_transmutation}
\end{align}
with $(r_3,r_4)\not=(0,0)$ being the $SU(N)$ adjoint color label. Thus,  for any finite action configurations, the gauge-field fluctuation should show the exponential decay to a flat connection controlled by the above mass gap as $|\bm{x}|\to \infty$, which looks like a collection of vortices on $\mathbb{R}^2$. 

We then consider the possible minimal-action configuration as the topological soliton for this setup. 
Since the gauge field $A_\mu$ exponentially converges to a pure gauge configuration as $|\bm{x}|\to \infty$, we can compactify $\mathbb{R}^2$ into $T^2$ with introducing a suitable 't~Hooft twist: 
\begin{equation}
    \mathbb{R}^2\times \underbrace{T^2}_{p/N\text{-twist}}\Rightarrow 
    \underbrace{T^2}_{-q/N\text{-twist}}\times \underbrace{T^2}_{p/N\text{-twist}}. 
\end{equation}
For such a twisted boundary condition, the topological charge $Q_{\mathrm{top}}=\frac{1}{8\pi^2}\int \tr(F\wedge F)$ is given by~\cite{vanBaal:1982ag}
\begin{equation}
    Q_{\mathrm{top}}\in \frac{pq}{N}+\mathbb{Z},  
\end{equation}
which has the $1/N$ fractional shift totally controlled by the boundary condition. 
Since the Bogomol'nyi bound of the Yang-Mills action $S=\frac{1}{g^2}\int \tr(F\wedge \star F)$ is $S_{\mathrm{YM}}\ge \frac{8\pi^2}{g^2}|Q_{\mathrm{top}}|$, the candidate of the minimal-action configuration lives on the sector $q$ with 
\begin{equation}
    qp\equiv \pm 1 \bmod N,
\end{equation}
so that 
\begin{equation}
    Q_{\mathrm{top}}=\pm \frac{1}{N},\quad S_{\mathrm{YM}}=\frac{8\pi^2}{Ng^2}. 
\end{equation}
We note that such a choice of $q$ is present as $\gcd(p,N)=1$. 
The label $q$ specifies the center phase rotation of the $2$d Wilson loop $W(C)$ for $C\subset \mathbb{R}^2$, $W(C)\to \rme^{\frac{2\pi\im}{N}q}W(C)$, when the loop surrounds the vortex configuration. 
This is nothing but the characterization of the center vortex~\cite{tHooft:1977nqb, Cornwall:1979hz, Nielsen:1979xu, Ambjorn:1980ms, DelDebbio:1996lih,  Faber:1997rp, DelDebbio:1998luz, Langfeld:1998cz, Kovacs:1998xm, Engelhardt:1999fd, deForcrand:1999our}, and the problem here is whether it is realized as the (anti-)self-dual Yang-Mills solution, $F=\pm \star F$, on twisted $\mathbb{R}^2\times T^2$. 

\subsection{Monopole-center vortex chain for  \texorpdfstring{$\ell_3\gg N\ell_4$}{l3>>N l4}}

Solving the self-dual Yang-Mills equation, $F=\pm \star F$, with the 't~Hooft twisted boundary condition on $\mathbb{R}^2\times T^2$ is a highly nontrivial task. 
We should note that Gonz\'alez-Arroyo and his collaborators have studied this problem numerically on the lattice in details~\cite{Gonzalez-Arroyo:1998hjb, Montero:1999by, Montero:2000pb, GarciaPerez:1989gt, GarciaPerez:1992fj}, 
and they confirmed that the self-duality bound is achieved by taking the continuum limit within a numerical error. 
Thus, there is little doubt on the presence of the self-dual center-vortex configuration itself, but we still would like to understand its analytic structures. 

On the other hand, the self-dual solutions on $\mathbb{R}^3 \times S^1$ with nontrivial holonomies are well understood analytically as the KvBLLY caloron~\cite{Lee:1997vp, Lee:1998bb, Lee:1998vu, Kraan:1998kp, Kraan:1998pm, Kraan:1998sn}, so it is natural to wonder if it helps the understanding for the self-dual center-vortex instanton. 
Following Refs.~\cite{Hayashi:2024yjc, Hayashi:2024psa, Hayashi:2025doq}, let us introduce the hierarchy for the two sizes of the torus: 
\begin{equation}
    \ell_3\gg N \ell_4. 
\end{equation}
As we have chosen the gauge $P_4=C$ being the diagonal clock matrix, the off-diagonal gluons acquire the gap of $\frac{2\pi}{N \ell_4}$, and they show the exponential decays much faster than those of diagonal gluons. 
Thus, the Abelianized description becomes effective for the $3$d dynamics at the length scale $\gg N\ell_4$ by simply neglecting the off-diagonal components of $A_\mu$. 
The $3$d Abelian gauge theory enjoys the monopole instantons, and the KvBLLY caloron has the fundamental constituents described as the self-dual counterparts of such monopoles~\cite{Lee:1997vp, Lee:1998bb, Lee:1998vu, Kraan:1998kp, Kraan:1998pm, Kraan:1998sn}, whose the magnetic charges are specified by the Affine simple roots, $\vec{\alpha}_n=\vec{\nu}_n-\vec{\nu}_{n+1}$, for $n=1,2,\ldots, N$. 
Here, $\vec{\nu}_n$ ($n=1,\ldots, N$) are the weight vectors for the defining representation of $SU(N)$. For the center-symmetric holonomy $P_4=C$, each monopole constituent carries $Q_{\mathrm{top}}=1/N$.

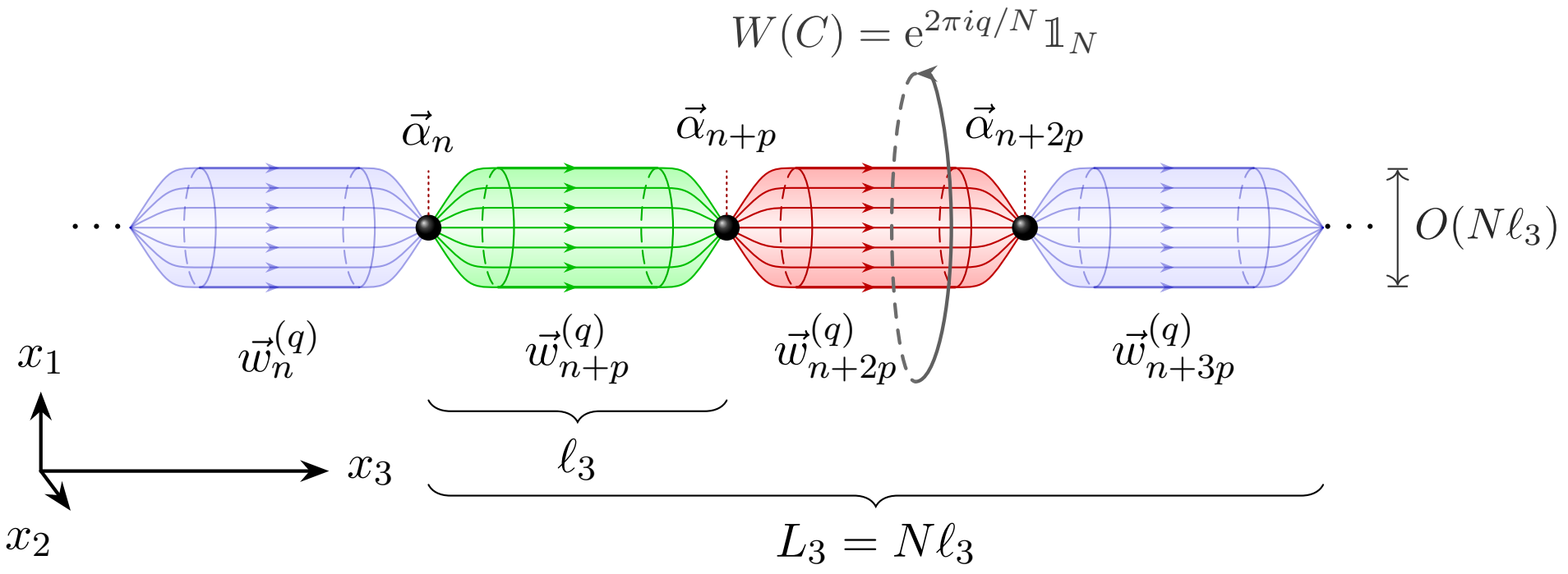
\begin{figure}[t]
    \centering
\begin{tikzpicture}[line cap=round,
  midarrow/.style={postaction={decorate},decoration={markings,
     mark=at position 0.5 with {\arrow{Stealth[length=3.2pt]}}}}, scale=0.8]
  \draw[-{Stealth},thick] (-3.9,-2.45) -- (-1,-2.45) node[right]{$x_3$};
  \draw[-{Stealth},thick] (-3.9,-2.45) -- (-3.9,-1.65) node[above]{$x_1$};
  \draw[-{Stealth},thick] (-3.9,-2.45) -- (-3.6,-2.85) node[below left]{$x_2$};
  \foreach \a/\col/\op in {-3/blue/0.35, 0/green/1, 3/red/1, 6/blue/0.35}{
    \pgfmathsetmacro{\b}{\a+3}
    \pgfmathsetmacro{\A}{\a+0.7}  \pgfmathsetmacro{\B}{\b-0.7}
    \pgfmathsetmacro{\ha}{\a+0.385}\pgfmathsetmacro{\hb}{\b-0.385}
    \begin{scope}[opacity=\op]
      \shade[top color=\col!30, bottom color=\col!30, middle color=\col!5]
        (\a,0) .. controls (\ha,0.6) .. (\A,0.6) -- (\B,0.6) .. controls (\hb,0.6) .. (\b,0)
               .. controls (\hb,-0.6) .. (\B,-0.6) -- (\A,-0.6) .. controls (\ha,-0.6) .. (\a,0);
      \draw[\col!70!black,semithick] (\A,0.6) -- (\B,0.6) (\A,-0.6) -- (\B,-0.6);
      \foreach \x in {\A,\B}{
        \draw[\col!70!black,thin,densely dashed] (\x,0.6)
          arc[start angle=90,end angle=270,x radius=0.16,y radius=0.6];
        \draw[\col!70!black,thin] (\x,-0.6)
          arc[start angle=-90,end angle=90,x radius=0.16,y radius=0.6];}
      \foreach \t in {-0.6,-0.4,-0.2,0,0.2,0.4,0.6}
        \draw[\col!75!black,thin,midarrow]
          (\a,0) .. controls (\ha,\t) .. (\A,\t) -- (\B,\t) .. controls (\hb,\t) .. (\b,0);
    \end{scope}}
  \node at (-3.3,0) {$\cdots$};
  \node at (9.3,0) {$\cdots$};
  \foreach \x/\lab in {0/{\vec\alpha_{n}}, 3/{\vec\alpha_{n+p}}, 6/{\vec\alpha_{n+2p}}}{
    \shade[ball color=black] (\x,0) circle (0.13);
    \draw[red!60!black,densely dotted] (\x,0.13) -- (\x,0.57);
    \node[black,above=1pt] at (\x,0.55) {$\lab$};}
  \foreach \x/\lab in {-1.5/{\vec w^{(q)}_{n}}, 1.5/{\vec w^{(q)}_{n+p}},
                       4.1/{\vec w^{(q)}_{n+2p}}, 7.5/{\vec w^{(q)}_{n+3p}}}
    \node at (\x,-1.22) {$\lab$};
  \draw[gray!75!black,thick,-{Stealth[length=5pt]}] (4.9,-1.55)
    arc[start angle=-90,end angle=90,x radius=0.36,y radius=1.55];
  \node[gray!45!black,font=\small,above=1pt] at (4.9,1.50) {$W(C)=\rme^{2\pi i q/N}\mathbbm{1}_N$};
  \draw[gray!85!black,thick,dashed] (4.9,1.55) arc[start angle=90,end angle=270,x radius=0.24,y radius=1.55];
  \draw[decorate,decoration={brace,amplitude=5pt,mirror,raise=2pt}]
    (0,-1.65) -- (3,-1.65) node[midway,below=7pt]{$\ell_3$};
  \draw[decorate,decoration={brace,amplitude=5pt,mirror,raise=2pt}]
    (0,-2.50) -- (9,-2.50) node[midway,below=7pt]{$L_3=N\ell_3$};
  \draw[|<->|,gray!40!black] (9.75,-0.6) -- (9.75,0.6)
    node[midway,right,font=\small,gray!30!black]{$O(N\ell_3)$};
\end{tikzpicture}
    \caption{Monopole–center vortex chain along $x_3$. 
    The magnetic flux $\vec{w}^{(q)}_{n+pk}$ is emitted by one monopole singularity $\vec{\alpha}_{n+p(k-1)}$ and absorbed by the next one $\vec{\alpha}_{n+pk}$ ($k\in \mathbb{Z}$). The size of each monopole core is about $N\ell_4 (\ll \ell_3)$. 
    The transverse size of the flux is $O(N\ell_3)$, and a Wilson loop encircling it acquires the Aharonov-Bohm phase, $\rme^{2\pi\im q/N}$. }
    \label{fig:monopole_vortex_chain}
\end{figure}

As we now consider the case $\ell_3\gg N\ell_4$, we can put a single monopole constituent on $\mathbb{R}^2\times (S^1)_{\ell_3}$ as the monopole core should look like a point, and let us put the $\vec{\alpha}_n$-monopole at $(\bm{x},x_3)=(\bm{0},0)$. 
The equation of motion becomes 
\begin{equation}
    \nabla^2 A_4^{\mathrm{Abel}}(\bm{x},x_3)=-2\pi \vec{\alpha}_n \delta^{(2)}(\bm{x})\delta(x_3),  
\end{equation}
where $A_4^{\mathrm{Abel}}=([A_4]_{mm})_{m=1,\ldots, N}$ is the Abelian component of $A_\mu$. 
Due to the self-duality, the color magnetic field is identical with the color electric field, so we skip the discussion on $A_{\mu=1,2,3}$ (see \cite{Hayashi:2024yjc, Hayashi:2024psa, Hayashi:2025doq} for the full discussion). 
This equation can be solved by the mirror-image method with paying attention to the boundary condition~\eqref{eq:algebraic_gauge}, $A_4(\bm{x},x_3+\ell_3)=S^{-p} A_4(\bm{x},x_3)S^{p}$, i.e. $[A_4]_{mm}(\bm x,x_3+\ell_3)=[A_4]_{m-p,m-p}(\bm x,x_3)$, so that the image of the $\vec\alpha_n$ source at $x_3=k\ell_3$ is $\vec\alpha_{n+pk}$.
By extending the domain of $x_3\in \mathbb{R}$, the equation of motion becomes 
\begin{equation}
    \nabla^2 A_4^{\mathrm{Abel}}(\bm{x},x_3)=-\sum_{k}2\pi \vec{\alpha}_{n+pk} \delta^{(2)}(\bm{x})\delta(x_3-k \ell_3), 
\end{equation}
which can be solved as (note that $A_4\to 0$ as $|\bm{x}|\to \infty$ in the algebraic gauge)\footnote{We note that Eq.~(2.34) of Ref.~\cite{Hayashi:2025doq} misses the overall factor $4$ while its Eqs.~(2.33), (2.35) are correct.}
\begin{align}
    A_4^{\mathrm{Abel}}&= \frac{1}{2}\sum_{k\in\mathbb{Z}}\vec{w}^{(q)}_{n+pk}
    \left(\frac{1}{|(\bm{x},x_3-k\ell_3)|}-\frac{1}{|(\bm{x},x_3-(k-1)\ell_3)|}\right) \notag \\
    &=\frac{2}{N\ell_3}\sum_{m=0}^{N-1}\vec{w}^{(q)}_{n+pm}\sum_{k'=1}^{\infty} K_0\left(\frac{2\pi}{N\ell_3}k'|\bm{x}|\right)\notag\\
    &\quad \times \left\{\cos \frac{2\pi k' (x_3-m \ell_3)}{N\ell_3}-\cos\frac{2\pi k' (x_3-(m-1)\ell_3)}{N\ell_3}\right\},
\end{align}
where 
\begin{equation}
    \vec{w}^{(q)}_{n}=\vec{\nu}_n+\vec{\nu}_{n+p}+\cdots+\vec{\nu}_{n+(q-1)p}.
\end{equation}
Here we take $qp\equiv +1\bmod N$, so that $\vec w^{(q)}_{n+pk}-\vec w^{(q)}_{n+p(k+1)}=\vec\nu_{n+pk}-\vec\nu_{n+pk+1}=\vec\alpha_{n+pk}$, and the first line reproduces the sources by summation by parts. For $qp=-1$ the same formula holds with $p\to-p$.
This clearly shows that the center-vortex configuration with the width $O(N\ell_3)$ comes out of the infinite alternating monopole chain as shown in Fig.~\ref{fig:monopole_vortex_chain}. 
By cutting out the segment $0\le x_3<\ell_3$, the Yang-Mills action and the topological charge are determined by the one of the KvBLLY monopole constituent, so $S_{\mathrm{YM}}=\frac{8\pi^2}{Ng^2}$ and $Q_{\mathrm{top}}=1/N$. 
The magnetic flux carried by the center vortex is $\vec{w}^{(q)}_{n}$, which belongs to the $q$-box anti-symmetric representation, and thus the Wilson loop acquires $\rme^{2\pi\im q/N}$ when it surrounds this vortex.

According to Fig.~\ref{fig:monopole_vortex_chain}, the monopole serves as the junction of the center vortex to change its color-magnetic direction. 
Even beyond the weakly coupled regime, such junctions in the center-vortex model play the important role for the confinement dynamics~\cite{DelDebbio:1997ke, Ambjorn:1999ym, deForcrand:2000pg, Engelhardt:1999wr, Oxman:2018dzp, Junior:2019fty, Junior:2022bol}: 
For example, a nonzero instanton density requires non-oriented center-vortex worldsheets, i.e., monopole junctions ~\cite{Engelhardt:1999xw, Reinhardt:2001kf, Cornwall:1999xw}. 
Moreover, the importance of the monopole junctions for the mass gap is recently uncovered in the numerical simulation of the lattice $\mathbb{Z}_N$ gauge theory~\cite{Nguyen:2024ikq, Giansiracusa:2025hfj}, and then it has been proven that the center-vortex proliferation without monopoles necessarily leads to a gapless phase, such as the Coulomb phase~\cite{Hayashi:2026brp}.

\subsection{Doubly-periodic instanton in the far-field limit}

We now consider the torus with the extended period, $x_3\sim x_3 + L_3$ and $x_4\sim x_4+L_4$, with 
\begin{equation}
    L_3=N\ell_3, \quad L_4=\ell_4, 
\end{equation}
where the 't~Hooft twist is untwisted: $P_3  P_4=P_4 P_3$ for asymptotic holonomies. 
We again consider the case where $L_4$ is sufficiently small and the nontrivial holonomy $P_4$ abelianizes the gauge group at low-energies, $SU(N)\xrightarrow{\mathrm{Higgs}}U(1)^{N-1}$. 

Let us discuss the far-field behavior for the self-dual gauge field with $Q_{\mathrm{top}}=1$, which is given by the Abelian self-dual gauge fields with the point-like KvBLLY monopole singularities~\cite{Lee:1997vp, Lee:1998bb, Lee:1998vu, Kraan:1998kp, Kraan:1998pm, Kraan:1998sn}. We write the location of $n$-th monopole constituent as $(\bm{y}_n, y_{3,n})\in \mathbb{R}^2\times S^1$ for $n=1,\ldots, N$:  
\begin{equation}
    \nabla^2 A_4^{\mathrm{Abel}}(\bm{x},x_3)=-\sum_{n=1}^{N}2\pi \vec{\alpha}_n \delta^{(2)}(\bm{x}-\bm{y}_n) \delta(x_3-y_{3,n}) 
    \label{eq:Abelian_CoulombEq}
\end{equation}
with $x_3\in \mathbb{R}/(L_3\mathbb{Z})$. 
We can solve it with the mirror-image method, and the $m$-th color component is given by
\begin{align}
    &\vec{\nu}_m\cdot A_4^{\mathrm{Abel}}(\bm{x},x_3)\notag\\
    &= \frac{1}{2}\sum_{k\in \mathbb{Z}}\left\{\frac{1}{|(\bm{x}-\bm{y}_m,x_3-y_{3,m}-kL_3)|}-\frac{1}{|(\bm{x}-\bm{y}_{m-1},x_3-y_{3,m-1}-kL_3)|}\right\} \notag\\
    &=\frac{1}{L_3}\ln\frac{|\bm{x}-\bm{y}_{m-1}|}{|\bm{x}-\bm{y}_m|}
    +\frac{2}{L_3}\sum_{k'=1}^{\infty}\left[ K_0\left(\frac{2\pi k' |\bm{x}-\bm{y}_m|}{L_3}\right)\cos\frac{2\pi k'(x_3-y_{3,m})}{L_3}\right.\notag\\
    &~~~~~~~~~~~~~~~~~~~~~~~~~~~~\left.\qquad - K_0\left(\frac{2\pi k' |\bm{x}-\bm{y}_{m-1}|}{L_3}\right)\cos\frac{2\pi k'(x_3-y_{3,m-1})}{L_3} \right]. 
\label{eq:HT-Abel}
\end{align}
Only when $\bm{y}_m=\bm{y}_{m-1}$, the logarithmic term disappears and the field strength shows the exponential decay with $\exp(-2\pi |\bm{x}|/L_3)$ as $|\bm{x}|\to \infty$. When $\bm{y}_m\not=\bm{y}_{m-1}$, the field strength in the $m$-th component decays as $O(1/|\bm{x}|^2)$, so it has the non-zero dipole tail.


\section{ADHM construction for the doubly-periodic instanton}
\label{sec:farfield2d}

We have seen the hint that the self-dual fractional instanton on $\mathbb{R}^2\times T^2_{p/N\text{-twisted}}$ comes out of the periodic alternating array of the KvBLLY monopole constituents, but the previous work~\cite{Hayashi:2024yjc, Hayashi:2024psa, Hayashi:2025doq} is totally limited to the far-field Abelian components. 
Can one really show that it appears as the far-field behavior of the fully non-Abelian self-dual configuration? 

In this section, we apply the ADHM technique for the charge-$1$ instanton on $\mathbb{R}^2\times T^2_{\text{untwisted}}$, i.e. the doubly-periodic instanton. 
This reduces the problem to solve the $2$d Green function for the Abelian Hitchin system with $N$ impurities on the dual torus $\hat{T}^2$~\cite{Ford:2002pa, Ford:2003vi, Ford:2005sq}. 
By extending the work by Bruckmann and van Baal~\cite{Bruckmann:2002vy, Bruckmann:2003ag} on multi-caloron solutions to the current setup, we derive the alternating monopole-center vortex chain from the ADHM data explicitly in the next section. 
This section serves as the theoretical foundation for it: We discuss the regularization scheme associated with the $2$d delta-function impurities in the Abelian Hitchin system, and justify the way to reconstruct the gauge field of the doubly-periodic instanton with using the Green's function.

\subsection{ADHM data for the doubly-periodic instanton}

In this subsection, we explain how the ADHM technique~\cite{Atiyah:1978ri} can be applied for constructing the doubly-periodic instanton. For the review on ADHM, see \cite{Corrigan:1983sv}. 

Let us start from the ADHM data for $k$ instantons in $SU(N)$ on $\RR^4\ni (x_\mu)_{\mu=1,\ldots, 4}$. 
Using the Pauli matrices $\tau_{j=1,2,3}$, we define 
\begin{equation}
    \sigma_\mu=(\im\tau_j,\mathbbm{1}),\qquad \sbar_\mu=(-\im\tau_j,\mathbbm{1}), 
    \label{eq:sigma}
\end{equation}
so that $\sigma_4=\sigma_0=\mathbbm{1}_2$. Note that $\sigma_{1,2,3}$ in the following are not Pauli matrices themselves but the ones multiplied by the imaginary unit. 
The ADHM data consist of
\begin{itemize}
    \item $\lambda$: an $N\times2k$ matrix, a bifundamental $2$-component spinor of $SU(N)\times U(k)$;
    \item $B=B_\mu\sigma_\mu$: a $2k\times2k$ matrix, where $B_\mu$ are $k\times k$ hermitian matrices.
\end{itemize}
We define $B(x):=B-x=\sigma_\mu(B_\mu-x_\mu\mathbbm{1}_{k})$ and introduce the Weyl operator as the $(N+2k)\times2k$ matrix,
\begin{equation}
\Delta(x):=\begin{pmatrix}\lambda\\ B(x)\end{pmatrix}.
\label{eq:Delta}
\end{equation}
The self-duality equation on the ADHM side can be expressed as
\begin{equation}
    \Delta(x)^\dagger\Delta(x)=
    \lambda^\dagger\lambda+B(x)^\dagger B(x)\overset{!}{=}\sigma_0\otimes f_x^{-1},
\label{eq:adhm-sd}
\end{equation}
where $f_x$ is a $k\times k$ matrix, called the Green function: \eqref{eq:adhm-sd} should be understood as three self-duality equations, $\frac{1}{2}\tr_2(\sigma_j[\lambda^\dagger \lambda+B(x)^\dagger B(x)])=0$ $(j=1,2,3)$, and the requirement for the invertiblity of $\frac{1}{2}\tr_2(\sigma_0[\lambda^\dagger \lambda+B(x)^\dagger B(x)])$, where $\tr_2$ is the trace for the spinor space.  

The gauge field $A_\mu(x)$ is constructed from the set of $N$ Weyl zero modes $v$, 
\begin{equation}
    \Delta(x)^\dagger v=0 ,
\label{eq:zeromodes}
\end{equation}
where $v$ is realized as the $(N+2k)\times N$ matrix, and the self-dual gauge field $A_\mu$ is then given by 
\begin{equation}
    A_\mu=-\im v^\dagger \partial_\mu v.
\label{eq:A}
\end{equation}
Using the Green function $f_x$, the self-dual gauge field $A_\mu$ turns out to be written as~\cite{Bruckmann:2002vy}
\begin{equation}
    A_\mu(x)=-\frac{\im}{2}\varphi(x)^{1/2}[\partial_\nu(\lambda [\overline{\eta}_{\mu\nu}\otimes f_x] \lambda^\dagger)] \varphi(x)^{1/2}-\frac{\im}{2}[\varphi(x)^{-1/2},\partial_\mu \varphi(x)^{1/2}], 
    \label{eq:GaugeField_GreenFunction}
\end{equation} 
where $\overline{\eta}_{\mu\nu}=\frac{1}{2}\overline{\sigma}_{[\mu}{\sigma}_{\nu]}$ is the anti-self-dual 't~Hooft eta symbol\footnote{We note that $F_{\mu\nu}$ is self-dual while the 't~Hooft symbol in this formula is the anti-self-dual one. The ADHM self-duality equation~\eqref{eq:adhm-sd} gives 
$vv^\dagger= \mathbbm{1}_{N+2k} - \Delta [\sigma_0\otimes f_x]\Delta^\dagger$, which shows
\begin{align}
    \im F
    &=\diff v^\dagger \wedge \diff v+ (v^\dagger \diff v)\wedge (v^\dagger \diff v)
    =\diff v^\dagger \wedge (1-vv^\dagger)\diff v \notag\\
    &= (\diff v^\dagger) \Delta \wedge f_x \Delta^\dagger \diff v= v^\dagger (\diff \Delta) \wedge f_x (\diff \Delta^\dagger) v. \notag
\end{align}
Since $\diff \Delta(x)=-\sigma_\mu \diff x_\mu$, this gives 
\begin{equation}
    F_{\mu\nu}=-2\im v^\dagger \eta_{\mu\nu} f_x v,  \notag
\end{equation}
with the self-dual 't~Hooft eta symbol, $\eta_{\mu\nu}=\frac{1}{2}\sigma_{[\mu}\sbar_{\nu]}$. 
} and 
\begin{equation}
    \varphi(x)^{-1}
    = \left[\mathbbm{1}_N+\lambda (B(x)^{\dagger}B(x))^{-1}\lambda^{\dagger}\right]^{-1} =  \mathbbm{1}_{N} - \lambda [\sigma_0\otimes f_x]\lambda^{\dagger}.
    \label{eq:def_varphi}
\end{equation}
In our convention, $A_\mu(x)$ is a hermitian matrix-valued gauge field. 

Next, let us bring this ADHM method for the doubly-periodic instanton, charge-$1$ instanton on $\mathbb{R}^2\times T^2$. 
Here, we write the periodicity of the torus as 
\begin{equation}
    x_3\sim x_3+L_3,\quad x_4\sim x_4+L_4, 
\end{equation}
and we distinguish it from the ones $\ell_{3,4}$ in Sec.~\ref{sec:review_centervortex_Abelian}. 
The reason is that we here work on the case without the 't~Hooft twist, and thus the asymptotic holonomies $P_{3}$ and $P_{4}$ commute with each other: $P_3P_4=P_4P_3$. Thus, we choose the algebraic gauge in which the transition functions $P_{3,4}$ are diagonal;
\begin{equation}
    P_a=\mathrm{diag}(\ee^{\im L_a\mu_{a,1}},\ldots, \ee^{\im L_a\mu_{a,N}})
    =\sum_{m=1}^N\ee^{\im L_a \mu_{a,m}}\,\Pi_m ,
\label{eq:P}
\end{equation}
where $\Pi_m$ is the projector onto the $m$-th color component.
The periodicity condition in the algebraic gauge is
\begin{equation}
    A_\mu(x+L_a \hat{e}_a)=P_a\,A_\mu(x)\,P_a^{-1}\qquad(a=3,4). 
\label{eq:A-period}
\end{equation}
To study the fractional-instanton situation in Sec.~\ref{sec:review_centervortex_Abelian}, we can unfold the twisted torus by extending the periodicity for the $3$rd direction; $x_3\sim x_3+N\ell_3$ and $x_4\sim x_4+\ell_4$. Therefore, we should identify
\begin{equation}
    L_3=N\ell_3, \quad L_4=\ell_4, 
\end{equation}
with specifying the holonomies as 
\begin{equation}
    P_3=(S^{-p})^{N}\propto \mathbbm{1}_N,\quad P_4=C.
\end{equation}
Center-symmetry transformation allows us to bring $P_3=\mathbbm{1}_N$, and, later, we will simply set $\mu_{3,m}=0$ and $\mu_{4,m}=\frac{2\pi}{NL_4}m$ (up to an irrelevant $m$-independent shift, needed for $\det P_4=1$) with $Q_{\mathrm{top}}=1$ to study the fractional-instanton setup in Sec.~\ref{sec:review_centervortex_Abelian}.
Let us note the holonomies alone do not ensure the unfolded instanton can be cut with the twisted torus: As we shall see in Sec.~\ref{sec:twist-cut}, this also requires a special choice of the remaining data $\zeta$.

To impose the boundary condition~\eqref{eq:A-period}, we should send $k\to \infty$ and require the lattice structure for the index as $I=(n_3,n_4)\in\mathbb{Z}\times \mathbb{Z}$~\cite{Kapustin:1998pb}, which is the straightforward extension for the KvBLLY caloron~\cite{Kraan:1998kp, Kraan:1998pm, Kraan:1998sn}.  
Define the $n_3$ and $n_4$ shift matrices acting on $X_{(n_3,n_4)}$ by
\begin{equation}
    (S_3X)_{(n_3,n_4)}=X_{(n_3-1,n_4)},\qquad (S_4X)_{(n_3,n_4)}=X_{(n_3,n_4-1)}.
\end{equation}
By requiring the ``periodicity condition'' for the ADHM data as
\begin{equation}
    S_a\lambda^\dagger P_a^{-1}=\lambda^\dagger,\qquad
    S_aBS_a^{-1}=B-L_a\sigma_a\qquad(a=3,4), 
\label{eq:adhm-period}
\end{equation}
we can find that $A_\mu(x)$ obeys \eqref{eq:A-period}.
This is solved as
\begin{equation}
    \lambda_{(n_3,n_4)}=P_3^{n_3}P_4^{n_4}\,\zeta,\qquad
    B_{I,J}=(n_4L_4\sigma_4+n_3L_3\sigma_3)\,\delta_{I,J}+\Ah_{I-J}, 
\label{eq:solved}
\end{equation}
where $\zeta:=\lambda_{(0,0)}$ and $\hat{A}_{I-J}:=B_{I-J,0}$, and these are the ADHM data for the doubly-periodic instanton with the given asymptotic holonomies, $P_{3,4}$.  

\subsection{Abelian Hitchin system with \texorpdfstring{$N$}{N} impurities}
\label{sec:hitchin}

To obtain the ADHM constraint for the ADHM data $(\zeta, \hat{A}_I)$ of the doubly-periodic instanton, it is useful to introduce the dual torus $\hat{T}^2\ni \xi=(\xi_3,\xi_4)$ with the periods $2\pi/L_{3,4}$ by the Fourier transform, $\hat{X}(\xi)=\sum_{n_3,n_4}\rme^{\im (n_3 L_3 \xi_3+n_4 L_4\xi_4)}X_{(n_3,n_4)}$, for $I=(n_3,n_4)\in \mathbb{Z}^2$~\cite{Kraan:1998kp, Kraan:1998pm, Kraan:1998sn, Kapustin:1998pb}. 
The detailed computation of the Fourier transform is done in Appendix~\ref{app:adhm-nahm}, and let us quote the result:
$B_{IJ}=(n_4 L_4 \sigma_4+n_3L_3\sigma_3)+\hat{A}_{I-J}$ becomes the $2$d Dirac operator on $\hat{T}^2$, 
\begin{equation}
    \hat B(\xi,\xi')=\Vh\,D\,\delta^{(2)}(\xi-\xi'),
\end{equation}
where $\hat{V}=4\pi^2/(L_3 L_4)$ is the volume of $\hat{T}^2$ and 
\begin{equation}
    D=-\im\sigma_4\partial_{\xi_4}-\im\sigma_3\partial_{\xi_3}+\Ah(\xi)=:\sigma_\mu D_\mu.
\label{eq:Dx}
\end{equation}
Here, $D_\mu$ is hermitian. 
$\lambda$ becomes the delta-function source specified by holonomies $P_{3,4}$, 
\begin{equation}
    \lamh^\dagger(\xi)
    =\Vh\sum_{m=1}^N\delta^{(2)}(\xi-\mu_m)\,\zeta^\dagger\Pi_m ,
\label{eq:lamhat}
\end{equation}
where 
\begin{equation}
    \mu_m=(\mu_{3,m}, \mu_{4,m})\in\hat T^2 ,\qquad m=1,\dots,N. 
\label{eq:ff-mum}
\end{equation}

The ADHM constraint \eqref{eq:adhm-sd}, $\Delta^\dagger\Delta=\lambda^\dagger\lambda+B(x)^\dagger B(x)=\sigma_0\otimes f_x^{-1}$, says that the $\sigma_l$ parts of $\Delta^\dagger\Delta$ vanish. Under the Fourier transformation, $B(x)\leftrightarrow D_x:=\sigma_\mu (D_\mu - x_\mu)$, and $B(x)^\dagger B(x)$ becomes
\begin{align}
    D_x^\dagger D_x&=\sbar_\mu\sigma_\nu (D_\mu-x_\mu)(D_\nu-x_\nu)\notag\\
    &= \sigma_0 (D_\mu-x_\mu)^2+ \frac{1}{2}\etab^l_{\mu\nu}\sigma_l [D_\mu, D_\nu]. 
\label{eq:DdD}
\end{align}
The other term $\lambda^\dagger \lambda$ becomes
\begin{equation}
    \widehat{\lambda^\dagger\lambda}(\xi)
    =\Vh\sum_{m=1}^N\delta^{(2)}(\xi-\mu_m)\,\zeta^\dagger\Pi_m\zeta .
\end{equation}
Following the convention of~\cite{Bruckmann:2002vy}, we parametrize the $2\times2$ matrix $\zeta^\dagger\Pi_m\zeta$ as
\begin{equation}
    \zeta^\dagger\Pi_m\zeta=:\frac1{2\pi}\bigl[\sigma_0 |\vec{\rho}_m|-\vec\tau\cdot\vec\rho_m\bigr].
    \label{eq:Srho}
\end{equation}
Since $\zeta^\dagger \Pi_m\zeta$ is rank-$1$ and semi-positive, the $\sigma_0$ component is totally determined by the other three components $\vec{\rho}_m$. Explicitly, $\zeta^\dagger\Pi_m\zeta=\chi_m\chi_m^\dagger$, where $\chi_m^\dagger$ is the $m$-th row of $\zeta$, so that $\rho^j_m=-\pi\chi_m^\dagger\tau_j\chi_m$ and $|\vec\rho_m|=\pi\chi_m^\dagger\chi_m$ (see \eqref{eq:rho}).

We can now obtain the Hitchin system with $N$ impurities from the ADHM constraint~\eqref{eq:adhm-sd}, $\tr[\sigma_l \Delta^\dagger \Delta]=0$:
\begin{align}
    \Fh_{4l}+\tfrac12\epsilon_{ljk}\Fh_{jk}=\frac{\Vh}{2\pi}\sum_m\rho^l_m\,\delta^{(2)}(\xi-\mu_m),
\label{eq:nahm-general}
\end{align}
where (setting $\partial_{\xi_{1,2}}=0$)
\begin{equation}
    \Fh_{\mu\nu}=\im[D_\mu,D_\nu]=\partial_{\xi_\mu}\Ah_\nu-\partial_{\xi_\nu}\Ah_\mu+\im[\Ah_\mu,\Ah_\nu]. 
\label{eq:Fhat}
\end{equation}
Since we consider the case with $Q_{\mathrm{top}}=1$, the dual $\hat{T}^2$ system becomes Abelian and the commutator term simply drops: $\hat{A}_3\diff \xi_3+\hat{A}_4\diff \xi_4$ is the $U(1)$ gauge field on $\hat{T}^2$ and $\hat{H}=\hat{A}_1+\im \hat{A}_2$ is the complex scalar field. Then, \eqref{eq:nahm-general} can be rewritten as
\begin{equation}
    \left\{
    \begin{aligned}
    &\Fh_{34}=-2\pi \sum_m\frac{\rho^3_m}{L_3L_4}\,\delta^{(2)}(\xi-\mu_m), \\
    &\frac{1}{2}(\partial_{\xi_3}-\im\partial_{\xi_4})\hat{H}=-\im\pi\sum_m\frac{\rho^1_m+\im\rho^2_m}{L_3L_4}\,\delta^{(2)}(\xi-\mu_m). 
    \end{aligned}
    \right.
    \label{eq:nahm}
\end{equation}
This is the Abelian Hitchin system that is dual to the $SU(N)$ doubly-periodic instanton~\cite{Kapustin:1998pb} (see also \cite{Ford:2000zt}). 
By integrating these equations on $\hat{T}^2$, we find the constraint, $\sum_m \vec{\rho}_m=0$, i.e. $\tr[\vec{\tau}\zeta^\dagger \zeta]=0$~\cite{Kraan:1998sn}, because $\Ah(\xi)=\sum_n\rme^{\im n\cdot(L\xi)}\Ah_n$ obtained from the ADHM data is a genuinely doubly-periodic function. 
In the following, we assume that $\kappa_m:=\rho^3_m/(L_3L_4)$ satisfies $0<|\kappa_m|<1$ for all $m$.\footnote{
A singular $U(1)$ gauge transformation on $\Td$ that winds by $+1$ around $\mu_m$ and by $-1$ around $\mu_{m'}$ maps the Nahm data to those with $(\kappa_m,\kappa_{m'})\to(\kappa_m+1,\kappa_{m'}-1)$, but it does not act on $\zeta$. 
The corresponding ADHM data have a different $\zeta$, since $|\chi_m|^2=|\vec\rho_m|/\pi$ changes, so the two sets of ADHM data are not related by a $U(k)\times SU(N)$ transformation. 
This non-equivalence under the singular $U(1)$ gauge transformation originates from the part where we drop $[\partial_{\xi_3},\partial_{\xi_4}]=0$ in the field strength $\hat{F}_{34}$. 
In the far-field computation, we shall see that the singular $U(1)$ gauge transformation shifts the $\alpha_m$-monopole constituents by $x_3\mapsto x_3+L_3$, and thus the gauge fields obtained from those ADHM data would coincide up to constituents' $L_3$-translation and the gauge translation despite the non-equivalence of the ADHM data. 
See also the discussion in Sec.~\ref{sec:twist-cut}. 
}

Formally, one can obtain the gauge field $A_\mu$ by using the Green function $f_x$ evaluated at impurities according to \eqref{eq:GaugeField_GreenFunction}. 
The Fourier transform of $f_x$ gives the kernel $\hat f_x(\xi,\xi')$ with\footnote{We note that $\hat f_x$ is the direct analogue of the Green function $f_x(z,z')$ of Bruckmann and van Baal~\cite{Bruckmann:2002vy}. 
By contrast, Ford and Pawlowski~\cite{Ford:2003vi} introduced the different Green function $\hat G$ for \eqref{eq:DdD}, i.e. $D_x^\dagger D_x\hat G=\hat V\delta^{(2)}$. Importantly, $\hat{G}_x$ is not proportional to $\sigma_0$ unlike $\hat f_x$. }
\begin{equation}
    \left[\sum_{\mu=1}^{4}(D_\mu-x_\mu)^2+\frac{\hat V}{2\pi}\sum_{m=1}^N |\vec{\rho}_m|\,\delta^{(2)}(\xi-\mu_m)\right]\,\hat f_x(\xi ,\xi')=\hat V\,\delta^{(2)}(\xi-\xi'), 
\label{eq:fhat}
\end{equation}
which is obtained by the Fourier transform of \eqref{eq:adhm-sd}. 
With the following formulas, 
\begin{align}
    \varphi^{-1}&=\mathbbm{1}_N-\sum_{m,m'=1}^N\hat f_x(\mu_m,\mu_{m'})\,\Pi_m\zeta\zeta^\dagger\Pi_{m'} ,
\label{eq:phi_f}\\
    \lambda[\sigma_j\otimes f_x]\lambda^\dagger&=\sum_{m,m'=1}^N\hat f_x(\mu_m,\mu_{m'})\,\Pi_m\zeta\,\sigma_j\,\zeta^\dagger\Pi_{m'} , 
\label{eq:eta_f}
\end{align}
Eq.~\eqref{eq:GaugeField_GreenFunction} determines $A_\mu(x)$ entirely in terms of the $N\times N$ matrix
\begin{equation}
    \left[\hat f_x(\mu_m,\mu_{m'})\right]_{m,m'=1,\ldots,N},
\label{eq:Fmatrix}
\end{equation}
whose entries are the values of the Green function between the impurity points. 
However, there is a subtlety related to the treatment of the $2$d $\delta$-function source, which shall be discussed below.

\subsection{The finite reconstruction of the self-dual gauge field}
\label{sec:renorm}

For the $1$d Nahm equation of calorons, the $\delta$-function terms in the analogue of \eqref{eq:fhat} only produce jump conditions, and $f_x(\mu_m,\mu_{m'})$ are manifestly finite~\cite{Bruckmann:2002vy}. On the two-dimensional dual torus this is no longer true, and \eqref{eq:phi_f}--\eqref{eq:eta_f} have to be read with care. The origin is the $k\to\infty$ limit on the ADHM side: every entry of $\lambda_{(n_3,n_4)}=P_3^{n_3}P_4^{n_4}\zeta$ has the same size, so $\lambda$ is not square-summable over the lattice $\mathbb Z^2$, which causes the ultraviolet divergence for $\hat{f}_x(\mu_m,\mu_{m'})$ and  $\varphi^{-1}$. 
Here, we will give an argument that such ultraviolet divergences cancel with each other for $A_\mu$ when using \eqref{eq:GaugeField_GreenFunction}, \eqref{eq:phi_f} and \eqref{eq:eta_f}, and the result becomes finite.

\subsubsection{Normalized Weyl zero modes for the Hitchin system}

For the case of $N=2$, this Hitchin system in the context of the doubly periodic instanton has been studied mathematically by Jardim~\cite{Jardim:1999dx, Jardim:1999mi}. 
Following Ford and Pawlowski~\cite{Ford:2002pa, Ford:2003vi, Ford:2005sq}, let us write 
\begin{equation}
    \kappa_m=\frac{\rho^3_m}{L_3L_4},\qquad \alpha_m=\frac{\rho^1_m+\im \rho^2_m}{L_3L_4}, 
\end{equation}
and we introduce $\gamma_m=|\vec{\rho}_m|/L_3L_4=\sqrt{\kappa_m^2+|\alpha_m|^2}$.
In Ref.~\cite{Jardim:1999dx}, Jardim has proven that the $SU(2)$ doubly-periodic instantons can be reconstructed from the admissible solutions of the Hitchin system with simple poles of the Higgs field, i.e. the case with $\alpha_m\not=0$.\footnote{
In Ref.~\cite{Jardim:2002tf}, Jardim claimed the bijection theorem between the
($1$-irreducible) $SU(N)$ doubly-periodic instanton and the admissible solutions
of the Hitchin system with $N$ impurities with the singular Higgs fields at most
having simple poles (see Theorem~$1$ and Theorem~$21$ of \cite{Jardim:2002tf} for
the rigorous statements), so the statement seems to include the case $\alpha_m=0$.
To our best knowledge, however, the argument for ``Hitchin solution $\to$
doubly-periodic instanton'' relies on the earlier work \cite{Jardim:1999dx},
where $\alpha_m\not=0$ is assumed.
It seems that the case $\alpha_m=0$ with $\kappa_m\notin\mathbb{Z}$ is included
in the later general treatment of the Nahm transform by
Mochizuki~\cite{Mochizuki:2013}, where a point at which the Higgs field is
regular but the parabolic weight is non-trivial is allowed. 
We note that, in Ref.~\cite[Theorem~3]{Jardim:2001cw}, the non-existence of doubly-periodic instantons was claimed to be proven for the $\alpha_m=0$ case, but we believe its proof hits an inconsistency (Concretely, the condition ``$b>0$ or $b'>0$'' for the degrees of line bundles at the singularities is derived from $b+b'=-1$, which leads to the non-existence claim, however the case $(b,b')=(-1,0)$ is missed there). 
} 
There, Jardim showed by the family index theorem that $\dim\ker D_x^\dagger=N=2$, i.e.\ the number of impurities, and then one can reconstruct the $SU(2)$ self-dual gauge field as 
\begin{equation}
    [A_\mu(x)]_{ij}=-\im \int_{\hat{T}^2}\diff^2 \xi\psi^\dagger_i(\xi;x) \partial_\mu \psi_j(\xi;x), 
\end{equation}
where $\psi_i(\xi;x)$ are $L^2$-normalizable Weyl zero modes of $D_x^\dagger$. 
This is also the route taken by Ford and Pawlowski~\cite{Ford:2002pa, Ford:2003vi, Ford:2005sq} for $\alpha_m=0$, while the full $x$-dependence is not explored there.

We interpret and reconstruct the Weyl zero modes of $D_x^\dagger$ from the ADHM data to make the connection with Jardim's work~\cite{Jardim:1999dx, Jardim:1999mi}. 
For the ADHM case, the zero modes $v$ for $\Delta(x)^\dagger v=0$ can be explicitly solved~\cite{Corrigan:1983sv} by putting
\begin{equation}
    v=\begin{pmatrix}-\mathbbm{1}_{N}\\ u\end{pmatrix}\varphi^{-1/2},
    \qquad u=\bigl(B(x)^\dagger\bigr)^{-1}\lambda^\dagger,
\label{eq:v}
\end{equation}
where $\varphi^{-1/2}>0$ makes the zero modes orthonormal, $v^\dagger v=\mathbbm{1}_N$. 
As we shall see explicitly in the far-field computation in Sec.~\ref{sec:far_field_limit}, the ultraviolet divergence causes $\varphi^{-1}\to 0$ for all the eigenvalues as we remove the ultraviolet cutoff (see Claim~\ref{ass:scaling}). Due to the orthonormal condition $v^\dagger v=\bm{1}_N$, this implies
\begin{equation}
    v\to 
    \begin{pmatrix}
        \bm{0}\\
        \psi(\xi;x)
    \end{pmatrix}
\end{equation}
with $u\varphi^{-1/2}\to \psi(\xi;x)=[\psi_i(\xi;x)]_{i=1,\ldots,N}$, as we remove the ultraviolet regulator.
Thus, $\varphi$ and also the Green function at impurities $\hat{f}_x(\mu_m,\mu_{m'})$ may suffer from the ultraviolet divergence, but the ADHM formula gives the normalized Weyl zero modes to give the finite self-dual gauge field $A_\mu(x)$.

\subsubsection{Point splitting and the renormalization prescription}
\label{sec:pointsplit}

Here, we give a concrete prescription to evaluate \eqref{eq:phi_f} and \eqref{eq:eta_f}, and to obtain the finite self-dual gauge field $A_\mu$ from \eqref{eq:GaugeField_GreenFunction}.
At points away from the impurities, $\hat f_x(\xi,\xi')$ is canonically defined by \eqref{eq:fhat} as the $2$d Green's function; the ultraviolet problem is only that the Green-function formula for $A_\mu$ require its values at the impurities.
Near the impurity $\mu_m$, the operator in \eqref{eq:fhat} is scale invariant at leading order: with $\varrho=|\xi-\mu_m|$, the Aharonov--Bohm flux $\kappa_m$ and the Higgs pole $|\hat H|^2\simeq|\alpha_m|^2/\varrho^2$ give the centrifugal term $[(\ell-\kappa_m)^2+|\alpha_m|^2]/\varrho^2$ in the sector of angular momentum $\ell$ around $\mu_m$.
In the $s$-wave sector $\ell=0$,  the two local behaviors are $\varrho^{\pm\gamma_m}$, and $\hat f_x$ takes the regular one, $\hat f_x(\xi,\xi')\propto\varrho^{\gamma_m}$ as $\xi\to\mu_m$ at fixed $\xi'$.
The values at the impurities therefore have to be defined by a limit with some renormalization prescription.

As a specific prescription, we split the points and average over directions: With some $\epsilon>0$, we define (identifying $\xi=\xi_3+\im \xi_4$)
\begin{equation}
    \hat f^\epsilon_x(\mu_m,\mu_{m'}):=\int_0^{2\pi}\frac{\diff\theta}{2\pi}\int_0^{2\pi}\frac{\diff\theta'}{2\pi}\;\hat f_x\big(\mu_m+\epsilon\,\rme^{\im \theta},\;\mu_{m'}+\epsilon\, \rme^{\im \theta'}\big),
\label{eq:pointsplit}
\end{equation}
i.e.\ each point source is replaced by the uniform distribution on the circle of radius $\epsilon$ around it. Let $\varphi_\epsilon$ and $M_{\epsilon,\mu\nu}:=\lambda[\etab_{\mu\nu}\otimes f_x^{\epsilon}]\lambda^\dagger$ be the matrices \eqref{eq:phi_f} and \eqref{eq:eta_f} with $\hat f_x(\mu_m,\mu_{m'})$ replaced by $\hat f^\epsilon_x(\mu_m,\mu_{m'})$. 
Due to the local nature of ultraviolet divergence, we may expect there exists a scaling behavior as $\epsilon\to 0$:  

\begin{claim}[Point-split renormalization and scaling ansatz]\label{ass:scaling}
Let $Z:=\mathrm{diag}(\epsilon^{2\gamma_1},\dots,\epsilon^{2\gamma_N})$ with the characteristic scale $\epsilon$ for the point-splitting regularization, then $\varphi_{\epsilon}^{-1}\to 0$ is characterized by the following scaling form (when $\gamma_m<1$ for all $m$),
\begin{equation}
    \tilde\varPhi^{-1}(x):=\lim_{\epsilon\to0}Z^{-1/2}\,\varphi_\epsilon^{-1}(x)\,Z^{-1/2},
\label{eq:Phit}
\end{equation}
where $\tilde\varPhi^{-1}$ is positive definite. Its entries
\begin{align}
    &\big[\tilde\varPhi^{-1}\big]_{mm'}=-\chi_m^\dagger\chi_{m'}\lim_{\epsilon\to0}\epsilon^{-\gamma_m-\gamma_{m'}}\,\hat f^\epsilon_x(\mu_m,\mu_{m'})\quad (m\neq m'),\notag\\
    &\big[\tilde\varPhi^{-1}\big]_{mm}=\lim_{\epsilon\to0}\epsilon^{-2\gamma_m}\left(1-\frac{|\vec\rho_m|}{\pi}\hat f^\epsilon_x(\mu_m,\mu_m)\right),
\label{eq:scaling}
\end{align}
and
\begin{equation}
    \tilde M_{\mu\nu}(x):=\lim_{\epsilon\to0}Z^{-1/2}\big(M_{\epsilon,\mu\nu}(x)-M^{(0)}_{\mu\nu}\big)Z^{-1/2}
\label{eq:Mt}
\end{equation}
exist together with their $x$-derivatives. Here\footnote{By \eqref{eq:phi_f}, \eqref{eq:eta_f} and \eqref{eq:Srho},
$
\big(M_{\epsilon,\mu\nu}\big)_{mm}=-\im\,\frac{\etab^l_{\mu\nu}\rho^l_m}{|\vec\rho_m|}\big(1-[\varphi_\epsilon^{-1}]_{mm}\big),
\label{eq:Mdiag}
$
so the subtraction in \eqref{eq:Mt} is $x$-independent and cancels the leading term of the diagonal.}, 
$M^{(0)}_{\mu\nu}=-\im\,\mathrm{diag}\Big(\frac{\etab^l_{\mu\nu}\rho^l_1}{|\vec\rho_1|},\dots,\frac{\etab^l_{\mu\nu}\rho^l_N}{|\vec\rho_N|}\Big)$.
\end{claim}

For $\alpha_m=0$, one can easily confirm Claim~\ref{ass:scaling} thanks to the results in \cite{Ford:2003vi}. There, we can take $\hat{H}=0$ and $\hat A={\star\diff\phi}$ (i.e.\ $\hat A_3=-\partial_{\xi_4}\phi$, $\hat A_4=\partial_{\xi_3}\phi$), with a single-valued real potential $\phi$ obeying $\nabla^2\phi=-2\pi\sum_m\kappa_m\delta^{(2)}(\xi-\mu_m)$, so that $\phi\simeq-\kappa_m\log|\xi-\mu_m|$ near $\mu_m$. The singularities at $\mu_m$ then factorize as $\exp(\pm \phi(\mu_m+\epsilon\rme^{\im\theta}))\sim\epsilon^{\mp\kappa_m}$ (see Ref.~\cite[Sec.~6]{Ford:2003vi}), which gives \eqref{eq:scaling} with $Z_m=\epsilon^{2|\kappa_m|}$. There, the condition $0<|\kappa_m|<1$ ($\Leftrightarrow \gamma_m<1$) is used for the $L^2$ normalization of the Weyl zero modes~\cite{Ford:2003vi}. 
For general $\alpha_m$, we do not have the proof but confirm the consistency in the far-field limit in the next section.

Accepting Claim~\ref{ass:scaling}, the gauge field is obtained as follows. At finite $\epsilon$, let $A^{(\epsilon)}_\mu$ denote \eqref{eq:GaugeField_GreenFunction} evaluated with $\varphi_\epsilon$ and $M_{\epsilon,\mu\nu}$, and we introduce $\tilde\varPhi_\epsilon:=Z^{1/2}\varphi_\epsilon Z^{1/2}$ and $\partial_\nu\tilde M_{\epsilon,\mu\nu}:=Z^{-1/2}\partial_\nu M_{\epsilon,\mu\nu}Z^{-1/2}$. $A^{(\epsilon)}_\mu$ is not self-dual at finite $\epsilon$, and only its limit is. For small $\epsilon$, $\varphi_\epsilon>0$ by the positivity of $\tilde\varPhi^{-1}$. Let
\begin{equation}
    g_\epsilon(x):=\varphi_\epsilon^{-1/2} Z^{-1/2} \tilde\varPhi_\epsilon^{1/2}.
\label{eq:gK}
\end{equation}
Since $Z^{-1/2}\tilde\varPhi_\epsilon Z^{-1/2}=\varphi_\epsilon$,
\begin{equation}
    g_\epsilon g_\epsilon^\dagger=\varphi_\epsilon^{-1/2}Z^{-1/2}\tilde\varPhi_\epsilon Z^{-1/2}\varphi_\epsilon^{-1/2}=\mathbbm{1}_N,\quad
    \det g_\epsilon=\frac{(\det\tilde\varPhi_\epsilon)^{1/2}}{(\det\varphi_\epsilon\,\det Z)^{1/2}}=1 ,
\end{equation}
so $g_\epsilon$ is an $SU(N)$-valued function.
The gauge transformation of $A^{(\epsilon)}_\mu$ by $g_\epsilon$ gives exactly
\begin{equation}
      \big(A^{(\epsilon)}_\mu\big)^{g_\epsilon}=-\frac{\im}{2}\tilde\varPhi_\epsilon^{1/2}[\partial_\nu\tilde M_{\epsilon,\mu\nu}]\tilde\varPhi_\epsilon^{1/2}-\frac{\im}{2}[\tilde\varPhi_\epsilon^{-1/2},\partial_\mu\tilde\varPhi_\epsilon^{1/2}].
\label{eq:A-ren}
\end{equation}
This identity holds at every $\epsilon$. By taking the limit $\epsilon\to0$, we obtain
\begin{equation}
    A'_\mu=-\frac{\im}{2}\tilde\varPhi^{1/2}\big[\partial_\nu\tilde M_{\mu\nu}\big]\tilde\varPhi^{1/2}-\frac{\im}{2}\big[\tilde\varPhi^{-1/2},\partial_\mu\tilde\varPhi^{1/2}\big],
\label{eq:A-renlim}
\end{equation}
which is written in terms of manifestly finite quantities.

$A^{(\epsilon)}_\mu$ converges also in the original gauge of \eqref{eq:GaugeField_GreenFunction}.
Let $\mathsf L_\epsilon:=Z^{1/2}\varphi_\epsilon^{1/2}$, then $\mathsf L_\epsilon (\mathsf L_\epsilon)^\dagger=\tilde\varPhi_\epsilon$.
$\mathsf L_\epsilon$ converges to the block-lower-triangular matrix $\mathsf L$ such that $\mathsf L\mathsf L^\dagger=\tilde\varPhi$, whose diagonal blocks with equal $\gamma_m$ are hermitian and positive. This is called the block Cholesky decomposition, and it is unique.\footnote{
Let us take the decreasing order for the indices $\gamma_m$. Since $\mathsf L_\epsilon\mathsf L_\epsilon^\dagger=\tilde\varPhi_\epsilon$ converges, $\mathsf L_\epsilon$ is bounded.
The hermiticity of $Z^{-1/2}\mathsf L_\epsilon=\varphi_\epsilon^{1/2}$ shows $(\mathsf L_\epsilon)_{m'm}^*=(Z_{m'}/Z_m)^{1/2}(\mathsf L_\epsilon)_{mm'}\to0$ for $\gamma_{m'}>\gamma_m$, so the limit is block lower triangular. Its diagonal blocks are limits of $Z_m^{1/2}$ times diagonal blocks of $\varphi_\epsilon^{1/2}>0$, hence hermitian and positive, and $|\det\mathsf L|^2=\det\tilde\varPhi>0$. Such a decomposition is unique, as each block is recursively obtained as $\mathsf L_{11}=\tilde\varPhi_{11}^{1/2}$, $\mathsf L_{21}=\tilde\varPhi_{21}\mathsf L_{11}^{-1}$, $\cdots$.
}
Hence,
\begin{equation}
    g_\epsilon\to g:=\mathsf L^{-1}\tilde\varPhi^{1/2}\in SU(N),\qquad A^{(\epsilon)}_\mu\to A_\mu:=(A'_\mu)^{g^{-1}},
\end{equation}
and
\begin{equation}
    A_\mu=-\frac{\im}{2}\mathsf L^\dagger [\partial_\nu \tilde{M}_{\mu\nu}]\mathsf L-\frac{\im}{2}\left[(\mathsf L^{-1}\partial_\mu \mathsf L)-(\mathsf L^{-1}\partial_\mu \mathsf L)^\dagger\right].
\end{equation}
In this sense, the formulas \eqref{eq:GaugeField_GreenFunction}, \eqref{eq:phi_f} and \eqref{eq:eta_f} hold with the values of $\hat f_x$ at the impurities defined by \eqref{eq:pointsplit} and $\epsilon\to0$.

\paragraph{Self-duality and periodicity.}
Let us discuss if the gauge field $A_\mu$ constructed above satisfies the requirement of the self-duality and the $T^2$ periodicity. 
Formally, the ADHM construction with the point-like source $\lambda$ gives a self-dual gauge field obeying the periodicity \eqref{eq:A-period}. 
The point splitting only regularizes divergent quantities whose combination in $A_\mu$ is finite, and thus it suffices to estimate how the regularization affects these properties. 
In the point-splitting regularization \eqref{eq:pointsplit}, the terms odd in $\theta$ or $\theta'$ drop out after the angular averages. 
The regularization gives an $x$-dependent correction of relative order $O(\epsilon^2)$, which becomes $O(\epsilon^{2-2\gamma_m})$ after multiplying by $Z^{-1/2}$ from both sides. 
The self-duality and the periodicity, which the most singular part satisfies as in the formal ADHM construction, can therefore be violated at finite $\epsilon$ by at most $O(\epsilon^{2-2\gamma_{\max}})$, i.e.
\begin{align}
    & ||F^{(\epsilon)}-\star F^{(\epsilon)}||=O(\epsilon^{2-2\max_m \gamma_m}),\notag\\
    & || A^{(\epsilon)}_\mu(x+L_a\hat{e}_a)-P_aA^{(\epsilon)}_\mu(x)P_{a}^{-1} || =O(\epsilon^{2-2\max_m \gamma_m}),
    \label{eq:finite_epsilon_violation}
\end{align}
and this vanishes as $\epsilon\to0$ for $\gamma_m<1$. 
A mathematically rigorous proof of these properties would be valuable, but it is beyond the scope of this paper.


\subsection{Constructing fractional instanton via 't~Hooft-twist cut}
\label{sec:twist-cut}

Let us now construct the fractional instanton on $\mathbb{R}^2\times T^2_{p/N\text{-twisted}}$ out of the doubly-periodic instanton.
In the algebraic gauge~\eqref{eq:algebraic_gauge}, a self-dual field obeys $A_\mu(x+\ell_3\hat e_3)=S^{-p}A_\mu(x)S^{p}$ and $A_\mu(x+\ell_4\hat e_4)=CA_\mu(x)C^{-1}$.
Regarded as a field on the $N$-fold cover $x_3\sim x_3+N\ell_3$, it obeys \eqref{eq:A-period} with $L_3=N\ell_3$, $L_4=\ell_4$, $P_3=(S^{-p})^N\propto\mathbbm{1}_N$, and $P_4=C$.
Thus, the fractional instanton with $Q_{\mathrm{top}}=1/N$ is a doubly-periodic instanton with $Q_{\mathrm{top}}=1$ and these holonomies, which in addition satisfies
\begin{equation}
    A_\mu(x+\ell_3\hat e_3)=S^{-p}A_\mu(x)S^{p}.
\label{eq:cut-condition}
\end{equation}
Conversely, any such doubly-periodic instanton descends to the twisted torus, and its fundamental domain $0\le x_3<\ell_3$ carries $Q_{\mathrm{top}}=1/N$ and $S_{\mathrm{YM}}=8\pi^2/(Ng^2)$.
We call this the 't~Hooft-twist cut of the doubly-periodic instanton.

Let us note that the construction of the Nahm transform on the twisted $T^4$ by passing to a covering torus and compensating a fractional shift on the dual torus by a phase and a change of basis of the Weyl zero modes was given by Gonzalez-Arroyo in Ref.~\cite{Gonzalez-Arroyo:1998nku} (see also Refs.~\cite{GarciaPerez:1999bc, Anber:2024uwl}); the twist cut here is its counterpart for $\mathbb{R}^2\times T^2$.

To translate \eqref{eq:cut-condition} into a condition on the ADHM data $(\zeta,\hat A)$, i.e., on the Hitchin data on $\hat T^2$, we first see how the two sides of \eqref{eq:cut-condition} act on the ADHM data.
The translation $A_\mu(x)\mapsto A_\mu(x+\ell_3\hat e_3)$ is generated by $B\mapsto B-\ell_3\sigma_3$, since $B(x)=B-x$.
In the dual torus, this is the constant shift
\begin{equation}
    \hat A_3(\xi)\mapsto \hat A_3(\xi)-\ell_3,
\label{eq:cut-translation}
\end{equation}
with $\hat A_4$, $\hat H$, and $\zeta$ unchanged.
Since $\ell_3=L_3/N$, this shifts the $U(1)$ holonomy along the $\xi_3$-cycle by $\omega=\rme^{2\pi \im/N}$, which is not a gauge transformation on $\hat T^2$.
The color permutation $A_\mu\mapsto S^{-p}A_\mu S^{p}$ is generated by $\lambda\mapsto S^{-p}\lambda$.
Using \eqref{eq:solved} with $P_3=\mathbbm{1}_N$ and $S^{-p}CS^{p}=\omega^{-p}C$, we find $(S^{-p}\lambda)_{(n_3,n_4)}=\omega^{-pn_4}C^{n_4}(S^{-p}\zeta)$.
Its phase $\omega^{-pn_4}$ can be removed by the $U(k)$ gauge transformation $\mathrm{diag}(\omega^{-pn_4})$, which multiplies $\hat A_{(n_3,n_4)}$ by $\omega^{-pn_4}=\rme^{-\im n_4L_4\cdot 2\pi p/(NL_4)}$, i.e., translates $\hat A(\xi)\mapsto\hat A\big(\xi-\tfrac{2\pi p}{NL_4}\hat e_4\big)$.
Since the rows of $S^{-p}\zeta$ are those of $\zeta$ shifted by $p$, the color rotation is equivalent to
\begin{equation}
    \big(\hat A(\xi),\hat H(\xi)\big)\mapsto\Big(\hat A\big(\xi-\tfrac{2\pi p}{NL_4}\hat e_4\big),\hat H\big(\xi-\tfrac{2\pi p}{NL_4}\hat e_4\big)\Big),\qquad \vec\rho_m\mapsto\vec\rho_{m-p}.
\label{eq:cut-color}
\end{equation}
This is the translation of the whole Hitchin configuration by $p$ impurities along $\xi_4$, which maps the impurity set onto itself, $\mu_{m-p}+\frac{2\pi p}{NL_4}\hat e_4=\mu_m$.
Thus, both sides of \eqref{eq:cut-condition} are fractional operations on $\hat T^2$, which gives 
\begin{equation}
    (\hat A_3-\ell_3,\hat A_4,\hat H;\vec\rho_m)\bigr|_{(\xi_3,\xi_4)}\sim (\hat A,\hat H;\vec\rho_{m-p})\bigr|_{(\xi_3, \xi_4-2\pi p/(NL_4))}, 
    \label{eq:cut-dual}
\end{equation}
i.e. they should be identical up to the $U(1)$ gauge transformation, which may be singular at impurities $\mu_m$. 
This leads to the following condition for the twist cut: 
\begin{equation}
    \alpha_m=0,\quad \text{and} \quad \kappa_m\equiv \frac{q}{N}\quad(\mathrm{mod}\ 1).
\label{eq:cut-data}
\end{equation}

To see $\alpha_m=0$, we note that $\hat H$ is neutral under the $U(1)$ gauge transformation, and thus its translation invariance by $\xi_4\mapsto \xi_4-\frac{2\pi p}{NL_4}$ gives $\alpha_{m-p}=\alpha_m$ for the residues. As $\gcd(p,N)=1$, all $\alpha_m$ are equal, and $\sum_m\alpha_m=0$ implies $\alpha_m=0$. Then $\hat H$ has no poles, and the second equation of \eqref{eq:nahm} gives the constant $\hat H=y_\perp$.

Let us next derive the condition for $\kappa_m$. For the gauge field, the gauge-invariant data on $\hat T^2\setminus\{\mu_m\}$ are the holonomies.
The holonomy around a small loop encircling $\mu_m$ (in counterclockwise) is $\rme^{2\pi\im\kappa_m}$, and the invariance under $\xi_4\mapsto \xi_4-\frac{2\pi p}{NL_4}$ gives $(\kappa:=)\kappa_{m-p}\equiv \kappa_m$ mod $1$. 
Also, the holonomy along the $\xi_3$-cycle at fixed $\xi_4$ gets multiplied by $\rme^{-2\pi\im\kappa_m}$ when $\xi_4$ crosses $\mu_{4,m}$ upwards.
The translation by $\frac{2\pi p}{NL_4}$ crosses $p$ impurities, which gives $\rme^{2\pi \im p \kappa}$, while this should be equivalent to $\hat{A}_3\mapsto \hat{A}_3-\ell_3$, which gives $\rme^{2\pi\im/N}$. 
Thus, we obtain $\rme^{2\pi\im p\kappa}=\rme^{2\pi \im/N}$.
Since $\sum_m\kappa_m=0$ gives $N\kappa\in\mathbb{Z}$, the $q$-th power of both sides gives $\rme^{2\pi \im \kappa}=\rme^{2\pi\im qp\kappa}=\rme^{2\pi \im q/N}$, and we find $\kappa\equiv q/N$ mod $1$.

The twist cut condition~\eqref{eq:cut-data} fixes all the impurity data, and thus the fractional instantons are parametrized only by the flat part of the Hitchin data:
\begin{itemize}
    \item the constant Higgs field $\hat H=y_\perp\in\mathbb{C}$, which is the position on $\mathbb{R}^2$;
    \item the holonomy of $\hat A$ along the $\xi_3$-cycle, which is the position along $x_3$;
    \item the holonomy of $\hat A$ along the $\xi_4$-cycle, which is the translation along $x_4$.
\end{itemize}
This gives the $4$ real moduli expected from the index theorem, $4NQ_{\mathrm{top}}=4$. 
Their global structure is discussed in Sec.~\ref{sec:summary}.
The twist cut data have $\alpha_m=0$, where the Hitchin system is solved explicitly by $\hat A=\star\diff\phi$ with $\nabla^2\phi=-2\pi\sum_m\kappa_m\delta^{(2)}(\xi-\mu_m)$. 
Claim~\ref{ass:scaling} is established there by the results of Ford and Pawlowski~\cite{Ford:2003vi}, and $\gamma_m=|\kappa_m|<1$.
Therefore, the construction of the fractional instanton in this subsection does not rely on the scaling ansatz beyond its proven case.

Let us now show that the data \eqref{eq:cut-data} indeed give a fractional instanton on $\mathbb{R}^2\times T^2_{p/N\text{-twisted}}$.
If \eqref{eq:cut-data} holds, the two sides of \eqref{eq:cut-dual} have the same constant Higgs field and the same holonomies on $\hat T^2\setminus\{\mu_m\}$.
They are therefore related by a $U(1)$ gauge transformation $\rme^{\im\theta(\xi)}$, which winds by the integer $\kappa_m-\kappa_{m-p}$ around $\mu_m$:
$D_{x+\ell_3\hat e_3}(\xi)=\rme^{\im\theta(\xi)}D_x\big(\xi-\tfrac{2\pi p}{NL_4}\hat e_4\big)\rme^{-\im\theta(\xi)}$ away from the impurities.
We should note that this is not a $U(k)$ transformation of the ADHM data, and thus we compare the Weyl zero modes instead, which gives the map $\psi(\xi)\mapsto\rme^{\im\theta(\xi)}\psi\big(\xi-\tfrac{2\pi p}{NL_4}\hat e_4\big)$.  
It maps the $L^2$-normalizable zero modes of $D_x^\dagger$ onto those of $D_{x+\ell_3\hat e_3}^\dagger$.
Both kernels are $N$-dimensional, since the fluxes on the two sides are only permuted and satisfy $0<|\kappa_m|<1$.
Hence the normalized zero modes are related as
\begin{equation}
    \psi(\xi;x+\ell_3\hat e_3)=\rme^{\im\theta(\xi)}\,\psi\big(\xi-\tfrac{2\pi p}{NL_4}\hat e_4;x\big)\,\Omega_3(x)^{-1}
\label{eq:cut-zeromode}
\end{equation}
with some $\Omega_3(x)\in SU(N)$.
By reconstructing $A_\mu$ from $\psi$, it obeys $A_\mu(x+\ell_3\hat e_3)=A_\mu^{\Omega_3^{-1}}(x)$. The phase $\rme^{\im\theta}$ drops out because $\theta$ does not depend on $x$, and the translation in $\xi$ drops out after integrating over $\hat T^2$. 

The twist is also read off from the zero modes. The ADHM frame obeys $\psi(\xi;x+\ell_4\hat e_4)=\rme^{\im\ell_4\xi_4}\psi(\xi;x)\,C^{-1}$ by \eqref{eq:adhm-period}. 
Computing $\psi(\xi;x+\ell_3\hat e_3+\ell_4\hat e_4)$ in the two orders, the translation $\xi_4\mapsto\xi_4-\frac{2\pi p}{NL_4}$ fails to commute with $\rme^{\im\ell_4\xi_4}$ by the phase $\rme^{-\im\ell_4\cdot 2\pi p/(NL_4)}=\omega^{-p}$, so that $\Omega_3(x+\ell_4\hat e_4)=\omega^{-p}\,C\,\Omega_3(x)\,C^{-1}$. Hence the transition functions $\Omega_3(x)$ and $\Omega_4=C$ obey $\Omega_3(x+\ell_4\hat e_4)\Omega_4=\omega^{-p}\Omega_4\Omega_3(x)$.
Therefore the quotient is a self-dual field on $\mathbb{R}^2\times T^2_{p/N\text{-twisted}}$ with $Q_{\mathrm{top}}=1/N$. 
Some gauge transformation can bring it to the algebraic gauge \eqref{eq:algebraic_gauge} with $P_3=S^{-p}$ and $P_4=C$, where \eqref{eq:cut-condition} holds exactly.

In Sec.~\ref{sec:far_field_limit}, we shall see that the data \eqref{eq:cut-data} put all the constituents at the common transverse position $y_\perp$. Consecutive constituents are separated by $L_3\kappa_m\equiv q\ell_3$ along $x_3$, which reproduces the monopole--center vortex chain of Sec.~\ref{sec:review_centervortex_Abelian}.

\section{Monopole-center vortex chain in the far-field limit}
\label{sec:far_field_limit}

In this section, we derive the Abelian gauge potential in the far-field limit studied in Sec.~\ref{sec:review_centervortex_Abelian}.
To understand the monopole-center vortex array of the doubly-periodic instanton, we need the $(\bm{x},x_3)$-dependence while we may drop the $x_4$ dependence with assuming the smallness of $L_4$. 
The logical structure basically follows the one by Bruckmann and van Baal~\cite{Bruckmann:2002vy, Bruckmann:2003ag}: Their Nahm parameter $z\in[0,1]$ is replaced by $\xi_4\in [0,2\pi/L_4]$ and their $k\times k$ matrices are replaced by operators on $L^2(S^1_{2\pi/L_3})$, i.e. the function space of $\xi_3$.

\subsection{Far-field limit, the strip temporal gauge, and Abelianization}

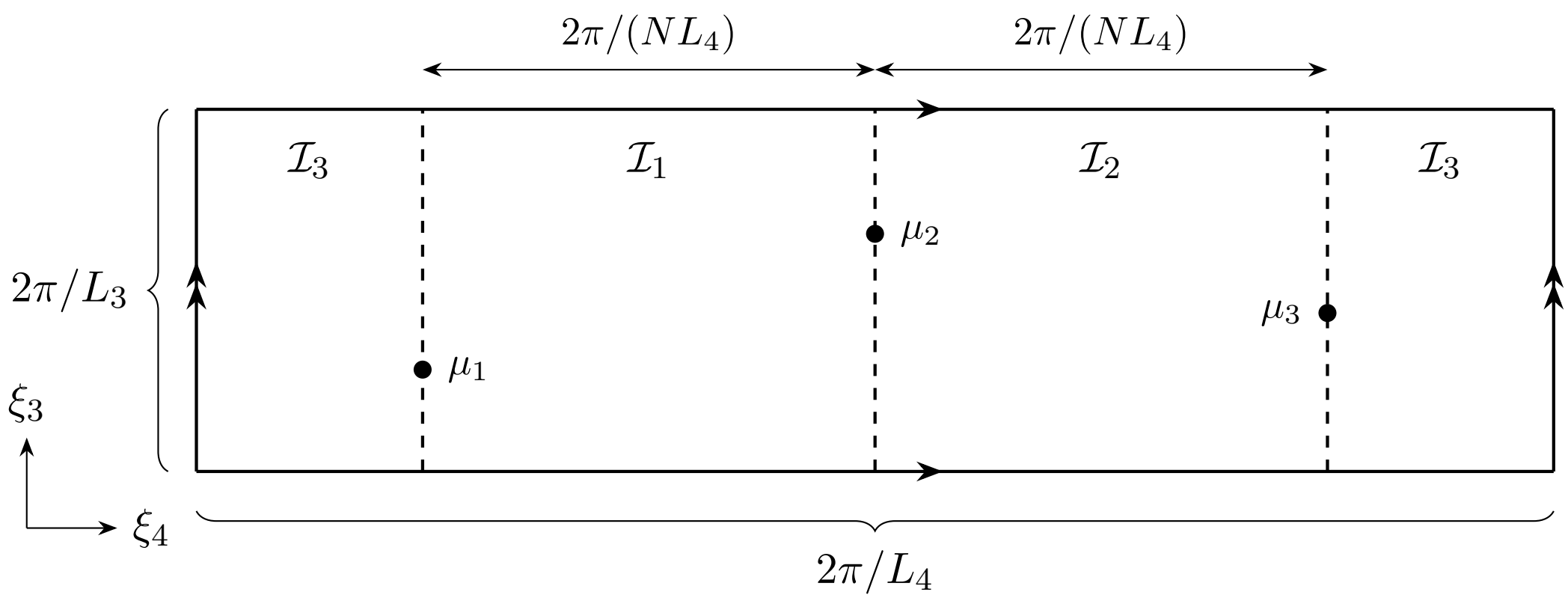
\begin{figure}[t]
\centering
\begin{tikzpicture}[
  >=Stealth,
  flux/.style={circle,fill=black,inner sep=1.6pt},
  ident1/.style={postaction={decorate},decoration={markings,mark=at position .55 with {\arrow{>}}}},
  ident2/.style={postaction={decorate},decoration={markings,mark=at position .52 with {\arrow{>}},mark=at position .58 with {\arrow{>}}}},
]
\def\W{12}   
\def\H{3.2}  
\draw[thick,ident1] (0,0) -- (\W,0);
\draw[thick,ident1] (0,\H) -- (\W,\H);
\draw[thick,ident2] (0,0) -- (0,\H);
\draw[thick,ident2] (\W,0) -- (\W,\H);
\foreach \x in {2,6,10} \draw[dashed,thick] (\x,0) -- (\x,\H);
\node[flux,label={[font=\small]right:$\mu_1$}] at (2,0.9) {};
\node[flux,label={[font=\small]right:$\mu_2$}] at (6,2.1) {};
\node[flux,label={[font=\small]left:$\mu_3$}] at (10,1.4) {};
\node at (4,2.75) {$\mathcal I_1$};
\node at (8,2.75) {$\mathcal I_2$};
\node at (11,2.75) {$\mathcal I_3$};
\node at (1,2.75) {$\mathcal I_3$};

\draw[decorate,decoration={brace,amplitude=5pt,mirror}] (0,-0.35) -- (\W,-0.35)
  node[midway,below=6pt] {$2\pi/L_4$};
\draw[decorate,decoration={brace,amplitude=5pt}] (-0.25,0) -- (-0.25,\H)
  node[midway,left=6pt] {$2\pi/L_3$};
\draw[<->] (2,\H+0.35) -- (6,\H+0.35) node[midway,above,font=\small] {$2\pi/(NL_4)$};
\draw[<->] (6,\H+0.35) -- (10,\H+0.35) node[midway,above,font=\small] {$2\pi/(NL_4)$};
\draw[->] (-1.5,-0.5) -- (-0.7,-0.5) node[right] {$\xi_4$};
\draw[->] (-1.5,-0.5) -- (-1.5,0.3) node[above] {$\xi_3$};
\end{tikzpicture}

\caption{Dual torus $\hat T^2$ for $N=3$ and the strip gauge. The point fluxes $\mu_m=(\mu_{3,m},\mu_{4,m})$ sit on the lines $\xi_4=\mu_{4,m}$ (dashed), spaced by $2\pi/(NL_4)$ for the twisted data. Strip $\mathcal I_m=\{\mu_{4,m}<\xi_4<\mu_{4,m+1}\}$ carries $\hat F_{34}=0$; there we set $\hat A_4=0$, and $\hat A_3$ and $\hat H$ have $\xi_3$-averages $y_{3,m}$ and $y_{\perp,m}$; the spectrum of $Y_{3,m}$ is $y_{3,m}+L_3\mathbb Z$. 
}
\label{fig:strip-gauge}
\end{figure}

Here, we explain the setup, where we perform the concrete calculation for constructing the self-dual gauge field $A_\mu$. 
We assume $P_4=C$, i.e. $\mu_{4,m}=\frac{2\pi m}{NL_4}$ up to a common shift for $m=1,\ldots, N$, and consider the case $NL_4\ll L_3$.\footnote{For the following computations, this assumption can be extended to the following setup without much effort: All the eigenvalues of $P_4$ are distinct, and $|\mu_{4,m}-\mu_{4, m+1}| \gg \frac{2\pi}{L_3}$. } 
For notational usefulness, we set $\mu_{4, N+1}=\mu_{4,1}+\frac{2\pi}{L_4}$.
We then divide the dual torus $\hat{T}^2$ into the $N$ strips (see Fig.~\ref{fig:strip-gauge}), 
\begin{equation}
    \mathcal{I}_m=\left\{(\xi_3,\xi_4)\in\hat{T}^2\,\Bigl|\, \mu_{4,m}<\xi_4<\mu_{4,m+1}\,\right\}. 
\label{eq:ff-strips}
\end{equation}

Let us discuss the behaviors of the $U(1)$ gauge field $\hat{A}_{a=3,4}$ and the Higgs field $\hat{H}$ inside each strip $\mathcal{I}_m$. 
While the Hitchin equation~\eqref{eq:nahm} can be solved with using the theta functions to obtain $\hat{A}_{a}$ and $\hat{H}$, let us extract the convenient information for the following analysis instead of giving the exact expression. 
For the Higgs field, we define 
\begin{equation}
    y_{\perp}(\xi_4)=y_{1}(\xi_4)+\im y_{2}(\xi_4):= \frac{L_3}{2\pi}\oint_{0}^{\frac{2\pi}{L_3}}\diff \xi_3 \hat{H}(\xi_3,\xi_4), 
\end{equation}
and then the Hitchin equation~\eqref{eq:nahm} tells
\begin{equation}
    \partial_{\xi_4} y_{\perp}(\xi_4)=L_3 \sum_m \alpha_m \delta(\xi_4-\mu_{4,m}). 
\end{equation}
Thus, inside each segment $\mathcal{I}_m$, $y_{\perp}(\xi_4)$ is constant:
\begin{equation}
    y_{\perp,m}=y_{1,m}+\im y_{2,m}:=y_{\perp}|_{\mathcal{I}_m}. 
\end{equation}
They are constrained by $\alpha_m$ as $y_{\perp,m}-y_{\perp,m-1}=L_3 \alpha_m$. 
Now, let us solve the Hitchin equation~\eqref{eq:nahm} for $\xi_4\approx \mu_{4,m}$, within the two strips $\mathcal{I}_{m-1}\cup \mathcal{I}_m$. 
Since $NL_4\ll L_3$, it is sufficient to solve it with the boundary condition $\hat{H}\to y_{\perp,m}$ as $\xi_4\to \infty$ and $\hat{H}\to y_{\perp,m-1}$ as $\xi_4\to -\infty$, which gives 
\begin{equation}
    \hat{H}(\xi_3,\xi_4)=\frac{y_{\perp,m}+y_{\perp,m-1}}{2}-\frac{\im (y_{\perp,m}-y_{\perp,m-1})}{2}\cot \frac{L_3\{(\xi_3-\mu_{3,m})-\im (\xi_4-\mu_{4,m})\}}{2}. 
\end{equation}
Thus, $\hat{H}$ converges to $y_{\perp,m}$ with the exponentially good accuracy with $O(\rme^{-L_3|\xi_4-\mu_{4,m}|})$ inside the strip $\mathcal{I}_m$.

For the $U(1)$ gauge field, let us choose the temporal gauge $\hat{A}_4=0$ inside each strip $\mathcal{I}_m$, and then $\hat{A}_3$ depends only on $\xi_3$ inside $\mathcal{I}_m$ as $\hat{F}_{34}=0$. 
As in the case of the Higgs field, we define 
\begin{equation}
    y_{3,m}:=\frac{L_3}{2\pi}\oint_{0}^{\frac{2\pi}{L_3}}\diff \xi_3\, \hat{A}_3|_{\mathcal{I}_m} , 
\end{equation}
and they are constrained by $\kappa_m$ as $y_{3,m}-y_{3,m-1}=L_3 \kappa_m$. 
Let us write 
\begin{equation}
    Y_{3,m}:= D_3|_{\mathcal{I}_m}=-\im \partial_{\xi_3}+\hat{A}_3|_{\mathcal{I}_m}, 
\end{equation}
and then its spectrum is given by 
\begin{equation}
    y_{3,m}+L_3 \mathbb{Z}. 
\end{equation}
Combining the two, the strip $\mathcal I_m$ carries the data ($\bm y_m:=(y_{1,m},y_{2,m})$)
\begin{align}
    &\vec y_{m,k}:=(\bm y_m,\,y_{3,m}+kL_3)\in\mathbb R^3,\qquad \notag\\
    &\vec y_{m,k}-\vec y_{m-1,k}=\frac{1}{L_4}\,\vec \rho_m .
\label{eq:ff-strip-data}
\end{align}
The second relation combines the jumps of $y_{\perp}$ and $y_3$, so that $|\vec y_{m,k}-\vec y_{m-1,k}|=\gamma_mL_3$.
At the end of the computation, we will find that $\vec y_{m,0}=(\bm{y}_m,y_{3,m})\in \mathbb{R}^2\times S^1_{L_3}$ indicates the location of the $m$-th monopole constituent inside the doubly-periodic instanton (see~\cite{Kraan:1998kp, Kraan:1998pm, Kraan:1998sn} for the single-caloron case).

We define the far-field limit by 
\begin{equation}
    r_{\min}:=\min_{m}|x_{\perp}-y_{\perp,m}|\gg NL_4 ,
\label{eq:ff-region}
\end{equation}
where $x_{\perp}=x_1+\im x_2$. We neglect contributions of $O(\rme^{-2\pi r_{\min}/(NL_4)})$ in the following, which leads to the Abelianization. 
In order to see it, let us evaluate the off-diagonal entries, $\hat{f}_x(\mu_m,\mu_{m+1})$, for the Green function. 
At distance $\gg1/L_3$ from the boundary of $\mathcal{I}_m$, $\hat H=y_{\perp,m}$ up to exponentially small terms, and the equation for the Green function (in the strip temporal gauge) becomes 
\begin{equation}
    \Bigl[-(\partial_{\xi_4}-\im x_4)^2 + \underbrace{(Y_{3,m}-x_3)^2 + |y_{\perp,m}-x_{\perp}|^2}_{\mathrm{mass}^2}\Bigr] \hat{f}_x(\xi,\xi')=\hat{V}\delta^{(2)}(\xi-\xi'). 
\end{equation}
Since $\mathrm{mass}^2\ge |x_\perp-y_{\perp,m}|^2$, we obtain
\begin{equation}
    |\hat{f}_x(\xi,\xi')| \lesssim \exp\left(-|\xi_4-\xi'_4|\, |x_{\perp}-y_{\perp,m}|\right).
\end{equation}
The magnetic field and the $\delta$-potential decrease it further and do not spoil this bound.
Near the boundary of $\mathcal{I}_m$, $\hat H$ is not constant: Close to each $\mu_m$, $\hat H\simeq-\im\alpha_m/(\bar \xi-\bar\mu_m)$ with $\xi=\xi_3+\im\xi_4$. 
As we take the limit $\xi\to \mu_m$ and $\xi'\to \mu_{m+1}$, the Aharonov-Bohm point-like flux with the $\delta$-function source term produces the renormalization factor as indicated in Claim~\ref{ass:scaling}, and we obtain the estimate 
\begin{equation}
    \hat{f}_x(\mu_m,\mu_{m+1})\sim Z_m^{1/2} Z_{m+1}^{1/2} \exp\left(-\frac{2\pi}{NL_4} |x_{\perp}-y_{\perp,m}|\right). 
\label{eq:ff-offbound}
\end{equation}
For non-adjacent $m,m'$ the path crosses several strips and the suppression is stronger. 
As a consequence, the off-diagonal entries of the renormalized quantities behave as
\begin{equation}
    [\tilde\varPhi]_{m,m'}, [\tilde M_{\mu\nu}]_{mm'} = O(\rme^{-2\pi r_{\min}/(NL_4)})
\end{equation}
when $m\not=m'$, which are negligible in the far-field limit.
Thus, the $N\times N$ matrices become diagonal matrices, which gives the Abelianization.

One can also easily understand this Abelianization from the viewpoint of Weyl zero modes. 
Each of $N$ Weyl zero modes, $\psi_m$, is localized around $\xi_4=\mu_{4,m}$ and shows the exponential decay for $|\xi_4-\mu_{4,m}|\gtrsim 1/|x_{\perp}-y_{\perp,m}|$ because it shares the same behaviors with the Green function, 
and thus the off-diagonal components, $\int \diff^2\xi \psi^\dagger_m \partial_{\mu}\psi_{n}$ with $m\not=n$, have the exponentially small overlaps of $O(\rme^{-2\pi r_{\mathrm{min}}/(NL_4)})$.

\subsection{Abelianized gauge field and the monopole--vortex chain}
\label{sec:ff-HT}

Due to the Abelianization, we can focus on computing the diagonal components of $A_\mu$, and the problem is now reduced to calculating $\hat{f}_x(\mu_m,\mu_m)$ in the far-field limit. 
We first introduce some notation to follow Bruckmann and van Baal~\cite{Bruckmann:2002vy, Bruckmann:2003ag}. 
Let us write the Nahm data in the strip temporal gauge as 
\begin{equation}
    \vec R(\xi_4;\vec x):=
    \begin{pmatrix}
        x_1-\mathrm{Re}\,\hat H \\  
        x_2-\mathrm{Im}\, \hat{H}\\ 
        x_3-Y_3
    \end{pmatrix}.
\label{eq:ff-YR}
\end{equation}
Its components are understood as the operator $L^2(S^1_{2\pi/L_3})\to L^2(S^1_{2\pi/L_3})$, and the Hitchin equation~\eqref{eq:nahm} becomes $\partial_{\xi_4}\vec{R}=\im \vec{R}\times \vec{R}$ for $\xi_4\not=\mu_m$. 
Introducing $|\delta_m\rangle$ by $\langle\xi_3|\delta_m\rangle=\delta(\xi_3-\mu_{3,m})$, 
we have the following jump condition of the Nahm data between $\mathcal{I}_{m-1}$ and $\mathcal{I}_m$ across the boundary $\xi_4=\mu_{4,m}$ for the strip temporal gauge,
\begin{equation}
    \begin{gathered}
    Y_{3,m}-Y_{3,m-1}=2\pi\kappa_m\,|\delta_m\rangle\langle\delta_m|,\\
    \hat H(\cdot,\mu_{4,m}+0^+)-\hat H(\cdot,\mu_{4,m}-0^+)=2\pi\alpha_m\,|\delta_m\rangle\langle\delta_m|, 
    \end{gathered}
\label{eq:ff-jump}
\end{equation}
using the matrix notation. Let us also define, with the unit vector $\hat n_m:=\vec\rho_m/|\vec\rho_m|=(\mathrm{Re}\,\alpha_m,\mathrm{Im}\,\alpha_m,\kappa_m)/\gamma_m$,
\begin{align}
    \mathcal S_m&:=2\pi\gamma_m|\delta_m\rangle\langle\delta_m|\notag\\
    &=-\hat n_m\cdot\big[\vec R(\mu_{4,m}+0^+)-\vec R(\mu_{4,m}-0^+)\big],
\label{eq:ff-Skappa}
\end{align}
which describes the impurity contribution: 
\begin{equation}
    \left[-\partial_{\xi_4}^2 + \vec{R}(\xi_4;\vec{x})^2+\sum_m \delta(\xi_4-\mu_{4,m}) \mathcal{S}_m\right]\hat{f}_x(\xi,\xi')= \hat{V}\delta^{(2)}(\xi-\xi'). 
\end{equation}
Here we have dropped $x_4$: it is removed by the gauge transformation $\rme^{\im x_4\xi_4}$, which does not affect $\hat f_x(\mu_m,\mu_m)$ in the far-field calculation.
In general setups, the source term is affected by the gauge transformation to the strip temporal gauge, but this $\mathcal{S}_m$ is unaffected since $|\delta_m\rangle \langle \delta_m|$ does not change under the $U(1)$ gauge transformation. 
Deep inside the strip $\mathcal I_m$, $\vec R(\xi_4,\vec{x})$ converges to
\begin{equation}
    \vec{R}_m=\begin{pmatrix}
        x_1-y_{1,m}\\  
        x_2-y_{2,m}\\ 
        x_3-Y_{3,m}
    \end{pmatrix},
\label{eq:ff-RP}
\end{equation}
and $R_m:=|\vec{R}_m|=\sqrt{\vec{R}_m\cdot \vec{R}_m}$ gives the mass in the bulk of $\mathcal{I}_m$.

Let us write $\langle \xi_3|\psi(\xi_4)\rangle:=\hat{f}_x(\xi,\mu_m)$. 
Since we consider the far-field limit, we may solve its differential equation with the boundary condition, $\psi(\xi_4)\to 0$, as $\xi_4-\mu_{4,m}\to \pm \infty$, which neglects the contribution from the reflection via other impurities at $\xi_4=\mu_{4,m\pm 1}$. 
Following \cite{Bruckmann:2002vy}, we introduce the solutions $R^\pm_m(\xi_4)$ of the Riccati equations on each $\mathcal{I}_m$,
\begin{equation}
        \big(R^\pm_m\big)^2\pm\partial_{\xi_4}R^\pm_m=\vec R\cdot\vec R,\qquad \xi_4\in\mathcal I_m ,
\label{eq:ff-riccati}
\end{equation}
and $R_m^{\pm}\approx |\vec R_m|$ when $\xi_4$ is deep inside $\mathcal{I}_m$. 
Then, since $\psi$ is continuous at $\xi_4=\mu_{4,m}$, we find 
\begin{equation}
    \psi(\xi_4)=
    \left\{
    \begin{array}{lc}
        \mathcal{P}\exp\left[-\int_{\mu_{4,m}}^{\xi_4}R^{-}_m(z)\diff z\right]\psi(\mu_{4,m}) 
        & \quad (\xi_4\in \mathcal{I}_m), \\
        \mathcal{P}\exp\left[+\int_{\mu_{4,m}}^{\xi_4}R^{+}_{m-1}(z)\diff z\right]\psi(\mu_{4,m}) & \qquad (\xi_4\in \mathcal{I}_{m-1}).
    \end{array} 
    \right.
\end{equation}
This shows $\partial_{\xi_4}\psi=-R^{-}_m\psi$ for $\xi_4\in \mathcal{I}_m$ and $\partial_{\xi_4}\psi=R^+_{m-1}\psi$ for $\xi_4\in \mathcal{I}_{m-1}$. 
Then, the jump condition at $\xi_4=\mu_{4,m}$ gives 
\begin{equation}
    \Sigma_m |\psi\rangle = \hat{V} |\delta_m\rangle,\qquad 
    \Sigma_m:=R_m^{-}\left(\mu_{4,m}\right) + R^{+}_{m-1}\left(\mu_{4,m}\right)+\mathcal{S}_m,
\end{equation}
and we find the far-field expression for $\hat{f}_x(\mu_m,\mu_m)$~\cite[Sec.4.1]{Bruckmann:2002vy},
\begin{equation}
    \hat f_x(\mu_m,\mu_m)=\langle \delta_m|\psi\rangle =  \Vh\,\langle\delta_m|\Sigma_m^{-1}|\delta_m\rangle. 
\label{eq:ff-Sigma}
\end{equation}
Using the matrix determinant lemma, we can rewrite it as 
\begin{align}
    [\varphi^{-1}]_{mm}
    &=
    1-\frac{|\vec\rho_m|}{\pi}\hat f_x(\mu_m,\mu_m)
    =\frac{\det[\Sigma_m-2\mathcal S_m]}{\det\Sigma_m}\notag\\
    &=\frac{\det[R^-_m(\mu_{4,m})+R^+_{m-1}(\mu_{4,m})-\mathcal S_m]}{\det[R^-_m(\mu_{4,m})+R^+_{m-1}(\mu_{4,m})+\mathcal S_m]}.
\label{eq:ff-detlemma}
\end{align}
Here, it is crucial that $\mathcal{S}_m\propto |\delta_m\rangle \langle \delta_m|$ is a rank-$1$ matrix.\footnote{For the case of multi-caloron, this rewriting is possible only for special cases such as the axially symmetric case~\cite{Bruckmann:2002vy}, which corresponds to $\alpha_m=0$. For the charge-$1$ doubly-periodic instanton, $\mathcal{S}_m$ is always rank-$1$, and we can apply it including the cases with $\alpha_m\not=0$. } 
As we have seen in Sec.~\ref{sec:pointsplit}, $\varphi$ gets the ultraviolet divergence, which corresponds to the determinant operation for the infinitely large matrix in this expression. 
Since such divergence cancels when we compute $A_\mu$, we perform the formal manipulations on the determinant formula and introduce a convenient regularization at the stage when we compute it concretely.

\paragraph{Evaluation of the determinant.}
For $\alpha_m=0$, the Nahm data are constant on each strip, so $R^-_m=R_m$, $R^+_{m-1}=R_{m-1}$, and \eqref{eq:ff-detlemma} can be evaluated as for the axially symmetric multi-caloron~\cite[Eqs.~(88)--(90)]{Bruckmann:2002vy}. For $\alpha_m\neq0$, $\hat H$ varies within $O(1/L_3)$ of the boundary, and $R^\pm$ at $\xi_4=\mu_{4,m}$ are not known in closed form. Nevertheless, the determinant ratio depends only on the bulk data:\footnote{The most nontrivial step to proceed the computation for nonzero $\alpha_m$'s is the way to evaluate the determinant in \eqref{eq:ff-detlemma}, and it is totally left open also for the case of multi-calorons in \cite{Bruckmann:2002vy}. 
The proof below was obtained by extending the strategy by Bruckmann and van Baal with the help of the large language model, Claude. 
}

\begin{proposition}[Determinant formula]\label{prop:ff-det}
For general $\alpha_m$,
\begin{equation}
    \frac{\det[R^-_m+R^+_{m-1}-\mathcal S_m]}{\det[R^-_m+R^+_{m-1}+\mathcal S_m]}\bigg|_{\xi_4=\mu_{4,m}}
    =\prod_{k\in\mathbb Z}\frac{|\vec x-\vec y_{m,k}|+\hat n_m\cdot(\vec x-\vec y_{m,k})}{|\vec x-\vec y_{m-1,k}|+\hat n_m\cdot(\vec x-\vec y_{m-1,k})}.
\label{eq:ff-firstorder}
\end{equation}
\end{proposition}

\begin{proof}
Near $\xi_4=\mu_{4,m}$, $\im D_x$ reduces to $\partial_{\xi_4}+\vec\tau\cdot\vec R$ by \eqref{eq:Dx}, and the ADHM equation becomes
\begin{equation}
    (\partial_{\xi_4}+\vec\tau\cdot\vec R)^\dagger(\partial_{\xi_4}+\vec\tau\cdot\vec R)=\big(-\partial_{\xi_4}^2+\vec R\cdot\vec R\big)\otimes\mathbbm 1_2\qquad(\xi_4\neq\mu_{4,m}),
\label{eq:ff-calD}
\end{equation}
which follows from $\partial_{\xi_4}\vec R=\im\vec R\times\vec R$. 
By \eqref{eq:ff-jump} and \eqref{eq:ff-Skappa}, $\vec R$ jumps by $-\hat n_m\mathcal S_m$ at $\xi_4=\mu_{4,m}$: the jump points along $\hat n_m$, and its size is the source strength. In the spin basis in which $\hat n_m\cdot\vec\tau$ is diagonal, we write
\[
\vec\tau\cdot\vec R=\begin{pmatrix} a&b^\dagger\\ b&-a\end{pmatrix},\qquad a:=\hat n_m\cdot\vec R,\quad b:=(\hat e_1+\im\hat e_2)\cdot\vec R,
\]
where $(\hat e_1,\hat e_2,\hat n_m)$ is a right-handed orthonormal frame. Then $b$ is continuous at $\mu_{4,m}$, and $a(\mu_{4,m}-0^+)-a(\mu_{4,m}+0^+)=\mathcal S_m$.

Let $\Psi_+$ be a $2$-component spinor that solves $(\partial_{\xi_4}+\vec\tau\cdot\vec R)\Psi_{+}=0$ and decays into $\mathcal I_m$. By \eqref{eq:ff-calD}, both spin components decay, so $\partial_{\xi_4}\Psi=-R^-_m\Psi$, and hence $(\vec\tau\cdot\vec R-R^-_m)\Psi_{+}=0$. Writing these solutions as $\Psi_{+}= \begin{pmatrix}
    \psi_{+}\\
    \Gamma_+ \psi_{+}
\end{pmatrix}$, the two rows give
\[
R^-_m-a=b^\dagger\Gamma_+,\qquad R^-_m+a=b\,\Gamma_+^{-1}.
\]
Likewise, the solutions $\Psi_{-}$ of the same spinor equation but decaying into $\mathcal I_{m-1}$ satisfy $(\vec\tau\cdot\vec R+R^+_{m-1})\Psi_{-}=0$. 
By setting $\Psi_{-}= \begin{pmatrix}
    \psi_{-}\\
    \Gamma_{-}\psi_{-}
\end{pmatrix}$, we find $R^+_{m-1}+a=-b^\dagger\Gamma_-$ and $R^+_{m-1}-a=-b\,\Gamma_-^{-1}$. 
At $\xi_4=\mu_{4,m}$ we use $a(\mu_{4,m}+0^+)$ in the first pair and $a(\mu_{4,m}-0^+)$ in the second, and obtain
\begin{equation}
    R^-_m+R^+_{m-1}+\mathcal S_m=b^\dagger(\Gamma_+-\Gamma_-),\quad
R^-_m+R^+_{m-1}-\mathcal S_m=-b\,\Gamma_+^{-1}(\Gamma_+-\Gamma_-)\,\Gamma_-^{-1}.
\end{equation}
The non-commuting factor $\Gamma_+-\Gamma_-$ cancels in the ratio of determinants. Since the ratio is positive,
\begin{equation}
\frac{\det[R^-_m+R^+_{m-1}-\mathcal S_m]}{\det[R^-_m+R^+_{m-1}+\mathcal S_m]}\bigg|_{\xi_4=\mu_{4,m}}=\frac{1}{|\det\Gamma_+\,\det\Gamma_-|}\bigg|_{\xi_4=\mu_{4,m}} .
\label{eq:ff-factor}
\end{equation}

Both spin components of a decaying solution in $\mathcal{I}_m$ are transported by the same operator $\mathcal P\exp(-\int R^-_m\,\diff\xi_4)$. 
Thus, $\Gamma_+(\xi_4)$ evolves by conjugation along $\mathcal I_m$, $\partial_{\xi_4}\Gamma_{+}=-[R_m^{-},\Gamma_{+}]$, and $\det\Gamma_+$ does not depend on $\xi_4$; the same holds for $\Gamma_-$ along $\mathcal I_{m-1}$. We may therefore evaluate them in the bulk, where $\vec R=\vec R_j$ is a function of $Y_{3,j}$ alone and $R^\mp=R_j$:
\begin{equation}
    \Gamma_+=(R_m+\hat n_m\cdot\vec R_m)^{-1}\,b_m,\qquad \Gamma_-=-(b_{m-1}^\dagger)^{-1}(R_{m-1}+\hat n_m\cdot\vec R_{m-1}),
\end{equation}
where $b_j$ is $b$ in the bulk of $\mathcal I_j$.\footnote{The operators $b_j$ are invertible unless $\vec x$ lies on a line through some $\vec y_{j,k}$ along $\hat n_m$; 
the intermediate steps are singular there, but the final result is not.} Since $\vec y_{m,k}-\vec y_{m-1,k}$ is parallel to $\hat n_m$ by \eqref{eq:ff-strip-data}, $b_m$ and $b_{m-1}$ have the same eigenvalues, in particular $|\det b_m|=|\det b_{m-1}|$, and hence the ratio becomes $\det(R_m+\hat n_m\cdot\vec R_m)/\det(R_{m-1}+\hat n_m\cdot\vec R_{m-1})$. On the eigenvector of $Y_{3,j}$ with eigenvalue $y_{3,j}+kL_3$, $R_j+\hat n_m\cdot\vec R_j=|\vec x-\vec y_{j,k}|+\hat n_m\cdot(\vec x-\vec y_{j,k})$, which gives \eqref{eq:ff-firstorder}. 
\end{proof}

\paragraph{Abelian gauge field and monopole chain.} 
We now have all the ingredients to evaluate the gauge field $A_\mu$ in the far-field limit. 
Let us set $\vec y_{m-1,k}(s):=\vec y_{m-1,k}+s\hat n_m$, then we obtain $\frac{\diff}{\diff s}\log\big[|\vec x-\vec y_{m-1,k}(s)|+\hat n_m\cdot(\vec x-\vec y_{m-1,k}(s))\big]=-1/|\vec x-\vec y_{m-1,k}(s)|$, and $\vec y_{m-1,k}(\gamma_mL_3)=\vec y_{m,k}$ by \eqref{eq:ff-strip-data}. Thus \eqref{eq:ff-detlemma} and \eqref{eq:ff-firstorder} give
\begin{equation}
-\log\,[\varphi^{-1}]_{mm}=\sum_{k}\int_{0}^{\gamma_m L_3}\frac{\diff s}{|\vec x-\vec y_{m-1,k}(s)|},
\label{eq:ff-rodsum}
\end{equation}
which is the Coulomb potential of a periodic array of uniformly charged rods. 
The $k$-th rod contributes $\simeq\gamma_m/|k|$ at large $|k|$, and  
thus, with the symmetric cutoff $|k|\le K$, 
the sum diverges as $2\gamma_m\log K$ and independently of $x$. 
This plays the role of $-\log Z_m$ in Claim~\ref{ass:scaling}, with $K\sim1/(\epsilon L_3)$: the symmetric cutoff, used here in place of the point splitting for simplicity, removes the same local singularity at $\mu_m$, so the finite parts should differ only by an $x$-independent factor.
Hence, we partially confirm the Claim~\ref{ass:scaling}: The diagonal entries $\tilde\varPhi_m$ of $\tilde\varPhi$ are given by 
\begin{equation}
\log\tilde\varPhi_m(x)=\lim_{K\to\infty}\Big[\sum_{|k|\le K}\int_{0}^{\gamma_m L_3}\frac{\diff s}{|\vec x-\vec y_{m-1,k}(s)|}-2\gamma_m\log K\Big]+\text{const}.
\label{eq:ff-rods}
\end{equation}
up to $O(\rme^{-2\pi r_{\min}/(NL_4)})$.
Only the gradients of \eqref{eq:ff-rods} enter the gauge field below, and they are absolutely convergent sums.

Up to $O(\rme^{-2\pi r_{\min}/(NL_4)})$, $\tilde\varPhi$ and $\tilde M_{\mu\nu}$ are diagonal. Then, the commutator in \eqref{eq:A-renlim} vanishes, and $\mathsf L=\tilde\varPhi^{1/2}$. 
By the footnote to \eqref{eq:Mt}, $[\tilde M_{\mu\nu}]_{mm}=\im\,\etab^l_{\mu\nu}\hat n^l_m\,\tilde\varPhi_m^{-1}$, and \eqref{eq:A-renlim} gives 
\begin{equation}
    [A_\mu]_{mm}=-\frac12\etab^l_{\mu\nu}\hat n^l_m\,\partial_\nu\log\tilde\varPhi_m. 
\end{equation}
With $\etab^l_{4j}=\delta_{lj}$, $\etab^l_{jk}=\epsilon_{ljk}$ and $\nabla=(\partial_1,\partial_2,\partial_3)$, we obtain
\begin{align}
    &[A_4]_{mm}=-\tfrac12\,\hat n_m\cdot\nabla\log\tilde\varPhi_m,\qquad\notag\\
    &\big([A_1]_{mm},[A_2]_{mm},[A_3]_{mm}\big)=\tfrac12\,\hat n_m\times\nabla\log\tilde\varPhi_m .
\label{eq:ff-Acomp}
\end{align}
Since $\log\tilde\varPhi_m$ is harmonic away from the rods,
\begin{equation}
    \nabla\times\vec A_{mm}=-\tfrac12(\hat n_m\cdot\nabla)\nabla\log\tilde\varPhi_m=\nabla[A_4]_{mm},
\end{equation}
and thus \eqref{eq:ff-Acomp} is an Abelian self-dual field. There is no $x_4$-dependence at this order. 

Shifting $\vec x$ along $\hat n_m$ only moves the endpoints of the rods, which gives
$\hat n_m\cdot\nabla\log\tilde\varPhi_m=\sum_{k}\big(|\vec x-\vec y_{m-1,k}|^{-1}-|\vec x-\vec y_{m,k}|^{-1}\big)$.
The impurity $\mu_m$ carries the color $m$, and $\sum_m\vec\rho_m=0$ gives $\vec y_{0,k}=\vec y_{N,k}$.
Summation by parts then gives
\begin{equation}
    \nabla^2A_4^{\mathrm{Abel}}=-2\pi\sum_{k\in\mathbb Z}\sum_{m=1}^N\vec{\alpha}_m\,\delta^{(3)}(\vec x-\vec y_{m,k}),
\label{eq:ff-sumparts}
\end{equation}
which is nothing but \eqref{eq:Abelian_CoulombEq}. 
Thus, in the far field, the doubly-periodic instanton is a periodic chain of the $N$ KvBLLY monopole constituents. The constituent $\vec\alpha_m$ sits at $(\bm y_m,y_{3,m})$ modulo $L_3$. Its transverse position is fixed by the Higgs residues, $\bm y_m-\bm y_{m-1}=L_3(\mathrm{Re}\,\alpha_m,\mathrm{Im}\,\alpha_m)$, and its longitudinal position by the fluxes, $y_{3,m}-y_{3,m-1}=L_3\kappa_m$.

\paragraph{Wilson loop on $\mathbb{R}^2$.}
Lastly, we compute the Wilson loop along a large circle $C_R=\{|\bm x|=R\}$ in $\mathbb{R}^2$ at fixed $(x_3,x_4)$. 
Since the off-diagonal components are $O(\rme^{-2\pi R/(NL_4)})$, the Wilson loop is determined by the diagonal components \eqref{eq:ff-Acomp}. 
For $|\bm{x}|=R\gg L_3, |\bm{y}_m|$, we can approximate \eqref{eq:ff-rods} as 
\begin{align}
    \log\tilde\varPhi_m
    &\simeq  \lim_{K\to\infty}\left[\gamma_m L_3 \sum_{k=-K}^{K} \frac{1}{\sqrt{|\bm{x}|^2+(kL_3)^2}}-2\gamma_m \log K\right]+\mathrm{const.}
    \notag\\
    &\simeq \lim_{K\to\infty}\left[\gamma_m \int_{-KL_3}^{KL_3} \frac{\diff z}{\sqrt{|\bm{x}|^2+z^2}}-2\gamma_m \log K\right]+\mathrm{const.}
    \notag\\
    &=-2\gamma_m\log |\bm x|+\text{const}.+O(|\bm x|^{-1}),
\end{align}
where the $x_3$-dependence is suppressed by $\rme^{-2\pi |\bm{x}|/L_3}$. 
Substituting this into \eqref{eq:ff-Acomp}, only the $x_3$-component of $\hat n_m$, i.e.\ $\gamma_m\hat n^3_m=\kappa_m$, contributes to
\begin{align}
    &\big([A_1]_{mm},[A_2]_{mm}\big)=\kappa_m\frac{(x_2,-x_1)}{|\bm x|^2}+O(|\bm x|^{-2}),\qquad \notag\\
    &\oint_{C_R}[A]_{mm}=-2\pi\kappa_m+O(R^{-1}).
\end{align}
The Wilson loop at $C_{R\to \infty}$ becomes
\begin{equation}
    W(C_\infty)=\lim_{R\to\infty}\mathcal{P}\exp\Big(-\im\oint_{C_R}A\Big)
    =\mathrm{diag}\big(\rme^{2\pi\im\kappa_1},\ldots,\rme^{2\pi\im\kappa_N}\big).
\label{eq:ff-wilson}
\end{equation}
For generic impurity data, $W(C_\infty)$ is not a center element of $SU(N)$: the doubly-periodic instanton carries the Abelian magnetic flux $2\pi\kappa_m$ in the $m$-th color through $\mathbb{R}^2$, and it is not a center vortex. 
For the data relevant to the 't~Hooft-twist cut, all the fluxes are equal, $\kappa_m\equiv q/N$ mod $1$, and \eqref{eq:ff-wilson} becomes the center element,
\begin{equation}
    W(C_\infty)=\rme^{2\pi\im q/N}\,\mathbbm{1}_N. 
\end{equation}
This reproduces the Wilson loop winding around the center vortex in Sec.~\ref{sec:review_centervortex_Abelian}, where $q$ serves as the twist on the compactified $\mathbb{R}^2$. 
In Sec.~\ref{sec:review_centervortex_Abelian}, this phase was obtained by solving the Abelianized equations for an assumed monopole chain. Here it follows from the full self-dual field, since the off-diagonal components vanish as $\rme^{-2\pi R/(NL_4)}$ and \eqref{eq:ff-wilson} is exact in the limit $R\to\infty$.

\section{Summary and Discussion}
\label{sec:summary}

\paragraph{Summary.}

In this paper, we have studied the $SU(N)$ doubly-periodic instanton on $\mathbb{R}^2\times T^2$ through the ADHM construction, and we have shown that its far field is the periodic chain of KvBLLY monopole constituents that underlies the monopole--center-vortex chain found previously within the Abelianized description~\cite{Hayashi:2024yjc, Hayashi:2024psa, Hayashi:2025doq}.
Following Kraan and van Baal~\cite{Kraan:1998kp, Kraan:1998pm, Kraan:1998sn} (see also \cite{Kapustin:1998pb, Ford:2000zt}), we took the formal $k\to\infty$ limit of the charge-$k$ ADHM data on $\mathbb{R}^4$ and required the infinite label to carry the $\mathbb{Z}^2$ lattice structure dual to $T^2$. 
After the Fourier transformation, the ADHM constraint becomes the Abelian Hitchin system on the dual torus $\hat T^2$ with $N$ impurities, \eqref{eq:nahm}. 
The positions $\mu_m$ of the impurities are fixed by the holonomies $P_{3,4}$, and their strengths are determined by the fluxes $\kappa_m$ and the residues $\alpha_m$ of the simple poles of the Higgs field.
Unlike the one-dimensional Nahm problem for calorons, the Green-function formula~\eqref{eq:GaugeField_GreenFunction} for $A_\mu$ requires the Green function at the impurities, \eqref{eq:phi_f}--\eqref{eq:eta_f}, where it is singular since they are points on the two-dimensional dual torus $\hat T^2$.
We regularized these values by point splitting, and then formulated a scaling ansatz, Claim~\ref{ass:scaling}, under which the rescaled matrices $\tilde\varPhi$ and $\tilde M_{\mu\nu}$ are finite. 
We then estimate the violation of self-duality and/or periodicity at finite $\epsilon$ as \eqref{eq:finite_epsilon_violation}, and thus the reconstructed gauge field $A_\mu$ becomes self-dual and doubly periodic as $\epsilon\to 0$.

We then analyzed the far-field regime $r_{\mathrm{min}}\gg NL_4$ defined by \eqref{eq:ff-region} with $P_4=C$, by extending the multi-caloron analysis of Bruckmann and van Baal~\cite{Bruckmann:2002vy, Bruckmann:2003ag}. 
The Abelianization in the far-field limit can be shown due to the suppression of the off-diagonal components as $\rme^{-2\pi r_{\min}/(NL_4)}$. 
The diagonal components are obtained in closed form~\eqref{eq:ff-Sigma}-\eqref{eq:ff-firstorder}, 
which gives the Coulomb potential of a periodic array of uniformly charged rods of length $\gamma_mL_3$ along $\vec \rho_m$. 
The resulting $A_4^{\rm Abel}$ solves the Abelian Coulomb equation~\eqref{eq:ff-sumparts} with the $N$ KvBLLY constituents, whose positions are fixed by the impurity data, $\bm y_m-\bm y_{m-1}=L_3(\mathrm{Re}\,\alpha_m,\mathrm{Im}\,\alpha_m)$ and $y_{3,m}-y_{3,m-1}=L_3\kappa_m$. 
For the data $\kappa_m\equiv q/N$ ($\bmod~ 1$) and $\alpha_m=0$, which unfold the 't~Hooft-twisted torus with $L_3=N\ell_3$ and $L_4=\ell_4$, 
we obtain the monopole-center vortex chain reviewed in Sec.~\ref{sec:review_centervortex_Abelian}, and its minimal segment gives the center-vortex fractional instanton on $\mathbb{R}^2\times T^2_{p/N\text{-twisted}}$ for $\ell_3\gg N\ell_4$.

\paragraph{Moduli space of the fractional instanton.}

Let us discuss the moduli space $\mathcal M_{1/N}$ of the center-vortex fractional instanton on $\mathbb{R}^2\times T^2_{p/N\text{-twisted}}$, together with its metric. 
The index theorem gives $\dim\mathcal M_{1/N}=4NQ_{\mathrm{top}}=4$. 
In the ADHM description, the condition for the twisted cut fixes all the data up to translations: the common position $(\bm y,y_{3,0})$ and the translation along $x_4$. 
The corresponding zero modes are $\delta_\nu A_\mu=F_{\nu\mu}$, i.e.\ translations combined with compensating gauge transformations, which satisfy the background gauge condition by the equation of motion. 
For a self-dual field, $\tr F_{\mu\lambda}F_{\nu\lambda}=\frac14\delta_{\mu\nu}\tr F_{\rho\sigma}F_{\rho\sigma}$ holds pointwise, and the metric induced by $\langle a,b\rangle=\frac{2}{g^2}\int\tr a_\mu b_\mu$ is flat and fixed by the action alone,
\begin{equation}
    g_{\mu\nu}=\frac{2}{g^2}\int\diff^4x\,\tr F_{\mu\lambda}F_{\nu\lambda}=(S_\mathrm{I}/N)\,\delta_{\mu\nu}, 
\end{equation}
where $S_\mathrm{I}=\frac{8\pi^2}{g^2}$ is the instanton action. 

Thus, the nontrivial part is to determine the global structure of $\mathcal{M}_{1/N}$. 
With fixed transition functions, a translation by one period is not a gauge redundancy: $x_a\to x_a+\ell_a$ maps $A\mapsto P_a AP_a^{-1}$, which affects the transition function as $P_3P_4P_3^{-1}=\rme^{-2\pi\im p/N}P_4$ for $x_3\mapsto x_3+\ell_3$ and $P_4P_3P_4^{-1}=\rme^{2\pi\im p/N}P_3$ for $x_4\mapsto x_4+\ell_4$. 
Along $x_3$, this is already visible in the Abelian far field of Sec.~\ref{sec:far_field_limit}, whose diagonal components are shifted by $P_3$; it corresponds to the range $y_{3,0}\in[0,N\ell_3)=[0,L_3)$ of the unfolded data. 
Along $x_4$, the diagonal far field is invariant, and the orbit closes after $N$ periods only because of the off-diagonal ($W$-boson) components in the cores, which must be present to remove the Dirac singularities of the Abelian chain. 
Therefore,
\begin{equation}
    \mathcal M_{1/N}\simeq\mathbb{R}^2\times T^2_{N\ell_3,\,N\ell_4},\qquad \mathrm{vol}\,\mathcal M_{1/N}=N^2\ell_3\ell_4\,\mathrm{Area}(\mathbb{R}^2), 
\end{equation}
which is the $N^2$-enlarged torus that appears in the color--momentum transmutation~\eqref{eq:colormomentum_transmutation}. 
This is the counterpart of the observation by Davies, Hollowood, Khoze, and Mattis~\cite{Davies:1999uw} for monopole-instantons on $\mathbb{R}^3\times S^1$: There, the $U(1)$ collective coordinate $\Omega\in[0,2\pi)$ corresponds, at the center-symmetric holonomy, to the $x_4$-translation range $X_4\in[0,2\pi/v)=[0,N\beta)$ rather than $[0,\beta)$, and this produces their weight $\Lambda^3/N$ per monopole. 
On the twisted $\mathbb{R}^2\times T^2$, the same phenomenon occurs in both compact directions, and the $N$ monopole types of $\mathbb{R}^3\times S^1$ become a single object related by $\ell_3$ translations~\cite{Hayashi:2024yjc}.

As an application of this structure, let us revisit the computation of the gluino condensate of $\mathcal N=1$ $SU(N)$ supersymmetric (SUSY) Yang--Mills theory. 
On the twisted $\mathbb{R}^2\times T^2$, the perturbative vacuum is gapped and has no moduli. 
On $\mathbb{R}^3\times S^1$, Davies, Hollowood, Khoze, and Mattis~\cite{Davies:1999uw} also evaluate $\langle\tr\lambda\lambda\rangle$ directly from the gluino zero modes of a single monopole. However, the holonomy and the dual photon are moduli there, and their vacuum values must first be fixed by the monopole-induced superpotential; Terashima's localization calculation~\cite{Terashima:2014yua} likewise obtains the condensate from the superpotential. 
Here, the 't~Hooft twist fixes the vacuum classically, so either approach reduces $\langle\tr\lambda\lambda\rangle$ to the single fractional instanton, and the nonperturbative determination of the vacuum is not required.

In the $\mathcal{N}=1$ SUSY localization~\cite{Terashima:2014yua}, the bosonic action is deformed by the nilpotent-SUSY-exact term $\supset t\int |F-\star F|^2$, and the path integral is evaluated at $t\to\infty$. There, the gauge field localizes on self-dual configurations, and the nonzero-mode functional determinants cancel. 
Following Ref.~\cite{Davies:1999uw}, the condensate is then given as follows:
\begin{align}
    \langle\tr\lambda\lambda(x)\rangle
    &=M_{\mathrm{PV}}^3\,\underbrace{\rme^{-S_\mathrm{I}/N+\im \theta/N} }_{\text{fractional instanton}} 
      \underbrace{\int_{\mathcal M_{1/N}}\frac{\diff^4X}{(2\pi)^2}\Big(\frac{g^2S_\mathrm{I}}{N}\Big)^2}_{\text{bosonic moduli}}
      \underbrace{\int\frac{\diff^2\chi}{2g^2S_\mathrm{I}/N}}_{\text{fermionic moduli}}\,\tr\lambda\lambda(x;X,\chi)
      \notag\\
    &=M_{\mathrm{PV}}^3\,\rme^{-S_\mathrm{I}/N+\im \theta/N}\,
     \frac{(g^2S_\mathrm{I}/N)^2}{(2\pi)^2}\,
      \times \frac{\int_{\mathbb{R}^2\times T^2_{N\ell_3,N\ell_4}}\diff^4X\int\diff^2\chi\,\tr\lambda\lambda(x;X,\chi)}{\int_{\mathbb{R}^2\times T^2_{\ell_3,\ell_4}}\diff^4X\int\diff^2\chi\,\tr\lambda\lambda(x;X,\chi)}
      \notag\\
    &=\frac{16\pi^2\Lambda^3\,\rme^{\im\theta/N}}{N^2}\times N^2, 
\end{align}
where the strong scale is $\Lambda^3\rme^{\im \theta/N}=M_{\mathrm{PV}}^3\rme^{-S_{\mathrm{I}}/N+\im \theta/N}$ with using the holomorphic gauge-coupling convention. 
The first line is the measure of Ref.~\cite[Eq.~(3.20)]{Davies:1999uw} written with the four translational collective coordinates $X_\mu$. 
The two gluino zero-modes are generated by supersymmetry, $\lambda_\alpha=\frac12\chi_\beta(\sigma_\mu\bar\sigma_\nu)_\alpha{}^\beta F_{\mu\nu}(x-X)$. 
Their Grassmann integral gives the gauge-invariant density $\int\diff^2\chi\,\tr\lambda\lambda(x;X,\chi) = \tr F_{\mu\nu}F_{\mu\nu}(x-X)$, 
where the normalization of the fermionic zero-mode is given by $\int_{\mathbb{R}^2\times T^2_{\ell_3,\ell_4}}\diff^4X\int\diff^2\chi\,\tr\lambda\lambda(x;X,\chi)=2g^2S_\mathrm{I}/N$ (see also~\cite[Eq.~(4.2)]{Davies:1999uw}). 
Since $\tr\lambda\lambda$ is invariant under the $\ell_{3,4}$ shifts of $X_{3,4}$, the $N^2$ center images in $\mathcal{M}_{1/N}\simeq \mathbb{R}^2\times T^2_{N\ell_3,N\ell_4}$ have identical densities: Each copy of the physical spacetime contributes only $16\pi^2\Lambda^3/N^2$, and thus the $N^2$-extended torus is essential for the correct normalization. 
We note that this value, originally obtained by the weak-coupling instanton analysis, has been tested by the large-$N$ lattice computation~\cite{Bonanno:2024bqg} based on the large-$N$ twisted Eguchi-Kawai volume reduction~\cite{GonzalezArroyo:1982hz, GonzalezArroyo:2010ss}, which finds agreement within the present uncertainties.

The phase $\rme^{\im\theta/N}$ is the semiclassical manifestation of the mixed 't~Hooft anomaly between the $\mathbb{Z}_{2N}$ discrete chiral symmetry and the $\mathbb{Z}_N$ $1$-form center symmetry~\cite{Gaiotto:2017yup, Shimizu:2017asf}. 
Since the twisted compactification preserves the anomaly~\cite{Tanizaki:2022ngt, Tanizaki:2017qhf, Dunne:2018hog, Yamazaki:2017dra}, the $N$-fold vacuum degeneracy persists when we take the periodic boundary condition and take the large-volume limit for $\mathbb{R}^2$. 
On the other hand, the path integral at fixed $\theta$ on $\mathbb{R}^2\times T^2$ selects one of the $N$ vacua for the free boundary condition at infinity on $\mathbb{R}^2$, and the others, $\langle\tr\lambda\lambda\rangle_k=16\pi^2\Lambda^3\rme^{\im(\theta+2\pi k)/N}$, are reached by $\theta\to\theta+2\pi k$.
In the above SUSY computation on  $\mathbb{R}^2\times T^2_{p/N\text{-twist}}$, the latter is chosen as the boundary condition for $\mathbb{R}^2$ and one of the chiral-broken vacua is implicitly chosen (see \cite{Misumi:2019dwq} for the analogous case of the multi-charge Schwinger model).

We would like to comment on the relation of our above computation with the recent discussion on the gluino condensate at the twisted $T^4$ by Anber and Poppitz~\cite{Anber:2022qsz, Anber:2023sjn, Anber:2024mco, Anber:2025vjo}. 
In their earlier works~~\cite{Anber:2022qsz, Anber:2023sjn}, the moduli space was taken with $N$-fold periods in all four directions, and the result was normalized by the Witten index, which gave $N\times16\pi^2\Lambda^3$ with the extra factor $N$ compared with the weak-coupling instanton result on $\mathbb{R}^4$. 
This confusion was resolved in their later work~\cite{Anber:2024mco} by noticing the mismatch of the natural normalizations between the path integral and the Hamiltonian interpretation, which suggests the $N$-fold extension along the time direction should not be counted when the normalization is taken by the Witten index. 
In Ref.~\cite{Anber:2025vjo}, they emphasized the importance of summing over twisted sectors to select a single superselection sector in the infinite volume limit, which was correctly taken but a bit implicit in earlier works,
and it has a natural counterpart here. 
In the $\mathbb{R}^2\times T^2$ computation, the finiteness of the action allows any numbers of the center-vortex fractional instantons on $\mathbb{R}^2$, while they are restricted to be in $N\mathbb{Z} + q$ when $\mathbb{R}^2$ is compactified on $T^2$ with the fixed twist $q$. 
In this regard, the path integral on $\mathbb{R}^2$ with the free boundary condition automatically sums over $q\in \mathbb{Z}_N$, and the choice of $\theta$ projects onto a single vacuum.

An important open question is the fluctuation determinant around the center-vortex fractional instanton, which controls the center-vortex fugacity in non-supersymmetric theories (see \cite{Diakonov:2004jn} for calorons). 
It determines the magnitude of the topological susceptibility and of the string tension in the dilute vortex gas of pure Yang--Mills theory~\cite{Tanizaki:2022ngt}, and that of the fractionalized Kobayashi--Maskawa--'t~Hooft vertex for the $\eta'$ field in QCD~\cite{Tanizaki:2022ngt, Hayashi:2024qkm}. 
So far, its $N$ dependence has only been estimated by a scaling argument~\cite[footnote~6]{Hayashi:2025doq}. 
Even if a closed expression cannot be derived, its parametric dependence on $N$ and $p$ should already provide important information, and we leave it for future studies.

\paragraph{D-brane interpretation.}
The fractional instanton turns out to have a simple D-brane picture, which extends the one used by Lee, Lu and Yi for the KvBLLY caloron~\cite{Lee:1997vp, Lee:1998bb}.
Let us consider $N$ D4-branes along $x_{1,2,3,4}$ that support the $U(N)$ gauge field $F^{U(N)}$, and we introduce D2-branes inside D4 that are dissolved and give the first Chern numbers of the $U(N)$ bundle; 
$n_{34}$ D2-branes along $x_{1,2}$ for $\int_{T^2}\tr F_{34}^{U(N)}=2\pi n_{34}$, and $n_{12}$ D2-branes along $x_{3,4}$ for $\int_{\mathbb{R}^2}\tr F_{12}^{U(N)}=2\pi n_{12}$. One may also add D0-branes, whose number is the second Chern character of the $U(N)$ bundle.
Since the latter is an integer, separating the $U(1)$ part gives
\begin{equation}
    \frac{1}{8\pi^2}\int \tr(F^{U(N)}\wedge F^{U(N)})= Q_{\mathrm{top}}+\frac{n_{12}n_{34}}{N}\in\mathbb{Z},
    \label{eq:brane-charge}
\end{equation}
so that the $SU(N)$ topological charge obeys $Q_{\mathrm{top}}\equiv -n_{12}n_{34}/N$ mod $1$.
The fractional charge is induced by the intersection of the two kinds of D2-branes~\cite{Guralnik:1997sy, Guralnik:1997th, Poppitz:2026gfa}.
Let us focus on $n_{12}n_{34}=-pq$ with $pq\equiv 1$ mod $N$ and $Q_{\mathrm{top}}=1/N$, which requires $(1-pq)/N$ D0-branes; none D0-branes are needed especially for $p=q=1$.

We now T-dualize along $x_4$ with the dual coordinate $\tilde{x}_4$, which brings $N$ D4-branes along $x_{1,2,3,4}$ and $p$ D2-branes along $x_{1,2}$ into $N$ D3-branes along $x_{1,2,3}$ and $p$ D3-branes along $x_{1,2,4}$.
These intersecting D3-branes become a single D3-brane winding the $(N,-p)$ cycle of the $(x_3,\tilde{x}_4)$ torus~\cite{Hashimoto:1997gm} because $\gcd(N,p)=1$.
At fixed $x_3$, it appears as $N$ sheets on the dual circle, equally spaced by $2\pi/(N\ell_4)$. 
Removing the $U(1)$ part of the tilt, this realizes the twisted boundary condition with $P_3=S^{-p}$ and $P_4=C$.

The D2-branes along $x_{3,4}$ become $q$ D1-strings along $x_3$, which are dissolved into $q$ of the $N$ sheets as the magnetic flux $F_{12}$ and give the center-vortex flux $2\pi\vec{w}^{(q)}_{n}$.
Since the sheets reconnect under $x_3\to x_3+\ell_3$ due to the $p$ D3-branes along $x_{1,2,4}$, the $q$ D1-strings cannot close on themselves.
In each period $\ell_3$, one D1-string must leave its sheet and climb to the adjacent sheet along $\tilde{x}_4$, and this D1-segment stretched between adjacent D3-branes is the BPS or KK monopole~\cite{Lee:1997vp}.
Its length, $1/N$ of the dual circle, gives $Q_{\mathrm{top}}=1/N$ and $S=8\pi^2/(Ng^2)$.
This gives the D-brane interpretation of the monopole--center vortex chain of Fig.~\ref{fig:monopole_vortex_chain}.

In this paper, we have used the ADHM construction for the doubly-periodic instanton with $Q_{\mathrm{top}}=1$, and the fractional instanton on the twisted $\mathbb{R}^2\times T^2$ has been obtained by cutting out its $1/N$ segment through constraints on the ADHM data.
It would be interesting to see whether the above picture allows one to work directly with the fractional instanton, e.g., by deriving its moduli space and ADHM-type data from the open strings in this brane configuration. 
Let us note that the supersymmetry of this configuration, in which the D1-brane meets the tilted D3-brane at an angle, is not trivial (see e.g. \cite{Berkooz:1996km, Bergshoeff:1997kr}) and requires further study. 

\acknowledgments

The authors thank Erich Poppitz and Antonio Gonz\'alez-Arroyo for reading the early draft and having the useful discussions. 
Y.H. thanks Takeshi Morita for discussions on the D-brane picture.
This work was completed during the long-term workshop at the Yukawa Institute for Theoretical Physics (YITP), ``New Frontiers in QCD 2026'' (YITP-T-26-01), and the discussion there was quite valuable. 
Discussions during RIKEN iTHEMS-YITP workshop: ``Resurgence, Matrix model, String field theory and Instantons (or Really Marvelous Sparkling Ideas)'' (YITP-W-26-11) were also useful in developing this work.
This work was partially supported by Japan Society for the Promotion of Science (JSPS)
Research Fellowship for Young Scientists Grant No. 23KJ1161 (Y.H.), by JSPS KAKENHI
Grant No. 26K07106 (Y.T.), and by Center for Gravitational Physics and Quantum Information (CGPQI) at Yukawa Institute for Theoretical Physics.

\appendix
\section{Fourier transform of the doubly-periodic ADHM data}
\label{app:adhm-nahm}

In this appendix, we derive the dual description on $\Td$ quoted in Sec.~\ref{sec:hitchin}, 
starting from the doubly-periodic ADHM data \eqref{eq:solved} with the ADHM index $I=(n_3,n_4)\in\ZZ^2$.

\subsection{Dictionary and the ADHM data on the dual torus}

The lattice index $I=(n_3,n_4)$ is Fourier transformed to the $\xi=(\xi_3,\xi_4)$ coordinate of the dual torus $\Td$, $\xi_a\sim\xi_a+2\pi/L_a$ $(a=3,4)$, whose volume is $\Vh=4\pi^2/(L_3L_4)$. 
Let us write $I\cdot(L\xi):=n_3L_3\xi_3+n_4L_4\xi_4$. 
A vector $X_I$ and a matrix $M_{IJ}$ in the ADHM index are represented as
\begin{equation}
    \hat X(\xi)=\sum_{I\in\ZZ^2}\ee^{\im I\cdot(L\xi)}X_I,\qquad
    \hat M(\xi,\xi')=\sum_{I,J\in\ZZ^2}\ee^{\im(I\cdot(L\xi)-J\cdot(L\xi'))}M_{IJ}.
    \label{eq:FT}
\end{equation}
One can easily check the following rules:
\begin{itemize}
    \item the unit matrix $\delta_{I,J}$ becomes $\Vh\,\delta^{(2)}(\xi-\xi')$;
    \item matrix products become $\widehat{MN}(\xi,\xi')=\Vh^{-1}\int_{\Td}\diff^2\xi''\,\hat M(\xi,\xi'')\hat N(\xi'',\xi')$, and $\widehat{MX}(\xi)=\Vh^{-1}\int_{\Td}\diff^2\xi'\,\hat M(\xi,\xi')\hat X(\xi')$;
    \item inner products become $\sum_I X_I^\dagger Y_I=\Vh^{-1}\int_{\Td}\diff^2\xi\,\hat X(\xi)^\dagger\hat Y(\xi)$;
    \item multiplication by $n_aL_a$ becomes the derivative, $\sum_I\ee^{\im I\cdot(L\xi)}\,n_aL_a\,X_I=-\im\partial_{\xi_a}\hat X(\xi)$.
\end{itemize}
In particular, an operator of the form $\hat M(\xi,\xi')=\Vh\,\mathcal{O}_\xi\,\delta^{(2)}(\xi-\xi')$, with $\mathcal{O}_\xi$ a differential operator on $\Td$, acts on $\hat X$ simply as $\mathcal{O}_\xi$, and the product of two such operators is $\Vh\,\mathcal{O}_\xi\mathcal{O}'_\xi\,\delta^{(2)}(\xi-\xi')$.

\paragraph{About $B(x)$.}
Inserting \eqref{eq:solved}, $B(x)_{IJ}=(n_4L_4\sigma_4+n_3L_3\sigma_3-x)\,\delta_{I,J}+\Ah_{I-J}$, into \eqref{eq:FT}, 
we obtain
\begin{align}
    \hat B(x)(\xi,\xi')
    &=\sum_{J}\ee^{\im J\cdot L(\xi-\xi')}\Bigl[n^J_4L_4\sigma_4+n^J_3L_3\sigma_3-x\Bigr]
    +\sum_J\ee^{\im J\cdot L(\xi-\xi')}\sum_{n\in\ZZ^2}\ee^{\im n\cdot(L\xi)}\Ah_n\notag\\
    &=\Vh\,\Bigl[-\im\sigma_4\partial_{\xi_4}-\im\sigma_3\partial_{\xi_3}+\Ah(\xi)-x\Bigr]\delta^{(2)}(\xi-\xi') \notag\\
    &=\Vh\,D_x\,\delta^{(2)}(\xi-\xi'),
    \label{eq:app-Bhat}
\end{align}
where $(n^J_3,n^J_4)=J$, $\Ah(\xi)=\sum_n\ee^{\im n\cdot(L\xi)}\Ah_n=\sigma_\mu\Ah_\mu(\xi)$, and $D_x=\sigma_\mu(D_\mu-x_\mu)$ with $D$ given in \eqref{eq:Dx}, i.e.
\begin{equation}
    D_a=-\im\partial_{\xi_a}+\Ah_a(\xi)\quad(a=3,4),\qquad D_{1,2}=\Ah_{1,2}(\xi).
    \label{eq:app-Dmu}
\end{equation}
Setting $x=0$ gives the formula quoted in Sec.~\ref{sec:hitchin}. The hermiticity of $B_\mu$ means $(\Ah_\mu)_n^\dagger=(\Ah_\mu)_{-n}$, so $\Ah_\mu(\xi)$ is hermitian; since $-\im\partial_{\xi_a}$ is self-adjoint on $\Td$, each $D_\mu$ is hermitian, and $B(x)^\dagger$ corresponds to $D_x^\dagger=\sbar_\mu(D_\mu-x_\mu)$.

\paragraph{About $\lambda$.}
By \eqref{eq:solved} and \eqref{eq:P}, $\lambda_I^\dagger=\zeta^\dagger(P_3^{n_3}P_4^{n_4})^\dagger=\sum_{m}\ee^{-\im I\cdot(L\mu_m)}\zeta^\dagger\Pi_m$ with $\mu_m=(\mu_{3,m},\mu_{4,m})$. Then, 
\begin{equation}
    \lamh^\dagger(\xi)=\sum_{m=1}^N\sum_{I\in\ZZ^2}\ee^{\im I\cdot L(\xi-\mu_m)}\,\zeta^\dagger\Pi_m
    =\Vh\sum_{m=1}^N\delta^{(2)}(\xi-\mu_m)\,\zeta^\dagger\Pi_m ,
\end{equation}
which is \eqref{eq:lamhat}: the holonomies $P_{3,4}$ fix the positions $\mu_m\in\Td$ of the impurities. Likewise, $\sum_J\ee^{-\im J\cdot(L\xi')}\lambda_J=\lamh^\dagger(\xi')^\dagger$, so that, using $\Pi_m\Pi_{m'}=\delta_{mm'}\Pi_m$,
\begin{equation}
    \widehat{\lambda^\dagger\lambda}(\xi,\xi')=\lamh^\dagger(\xi)\,\lamh^\dagger(\xi')^\dagger
    =\Vh\,\delta^{(2)}(\xi-\xi')\times\Vh\sum_{m=1}^N\delta^{(2)}(\xi-\mu_m)\,\zeta^\dagger\Pi_m\zeta .
    \label{eq:app-lamlam}
\end{equation}
Thus $\lambda^\dagger\lambda$ acts on $\Td$ as the multiplication by the localized source $\Vh\sum_m\delta^{(2)}(\xi-\mu_m)\,\zeta^\dagger\Pi_m\zeta$. Writing $\chi_m^\dagger$ for the $m$-th row of the $N\times 2$ matrix $\zeta$,
\begin{equation}
    \zeta=\begin{pmatrix}\chi_1^\dagger\\ \vdots\\ \chi_N^\dagger\end{pmatrix},\qquad
    \zeta^\dagger\Pi_m\zeta=\chi_m\chi_m^\dagger,\qquad
    \rho^j_m=-\pi\,\chi_m^\dagger\tau_j\chi_m,\qquad |\vec\rho_m|=\pi\,\chi_m^\dagger\chi_m ,
    \label{eq:rho}
\end{equation}
where $\vec\rho_m$ is defined by \eqref{eq:Srho}; the last relation follows from $\tr_2(\chi_m\chi_m^\dagger)=|\chi_m|^2$.

\subsection{The ADHM constraint as the Hitchin system with impurities}

By the product rule, $\Delta^\dagger\Delta=\lambda^\dagger\lambda+B(x)^\dagger B(x)$ becomes $\Vh\,\mathcal{K}_x\,\delta^{(2)}(\xi-\xi')$ with
\begin{equation}
    \mathcal{K}_x=D_x^\dagger D_x+\Vh\sum_{m=1}^N\delta^{(2)}(\xi-\mu_m)\,\zeta^\dagger\Pi_m\zeta .
\end{equation}
With $\sbar_\mu\sigma_\nu=\delta_{\mu\nu}\sigma_0+\etab^{\,l}_{\mu\nu}\sigma_l$, where $\etab^{\,l}_{4j}=-\etab^{\,l}_{j4}=\delta_{lj}$ and $\etab^{\,l}_{jk}=\epsilon_{ljk}$, and the hermiticity of $D_\mu$, we obtain \eqref{eq:DdD}, in which the constants $x_\mu$ drop out of the commutator. By \eqref{eq:Fhat} and the antisymmetry of $\etab^{\,l}_{\mu\nu}$,
\begin{equation}
    \tfrac12\etab^{\,l}_{\mu\nu}[D_\mu,D_\nu]=-\tfrac{\im}{2}\etab^{\,l}_{\mu\nu}\Fh_{\mu\nu}
    =-\im\Bigl(\Fh_{4l}+\tfrac12\epsilon_{ljk}\Fh_{jk}\Bigr).
\end{equation}
On the other hand, by \eqref{eq:Srho} and $\tau_l=-\im\sigma_l$, $\zeta^\dagger\Pi_m\zeta=\frac{1}{2\pi}\bigl(\sigma_0|\vec\rho_m|+\im\,\sigma_l\rho^l_m\bigr)$. Decomposing $\mathcal{K}_x=\sigma_0\,\mathcal{K}^{0}_x+\sigma_l\,\mathcal{K}^{l}_x$, we find
\begin{align}
    \mathcal{K}^{l}_x&=-\im\Bigl(\Fh_{4l}+\tfrac12\epsilon_{ljk}\Fh_{jk}\Bigr)+\frac{\im\Vh}{2\pi}\sum_{m=1}^N\rho^l_m\,\delta^{(2)}(\xi-\mu_m),
    \label{eq:app-Kl}\\
    \mathcal{K}^{0}_x&=\sum_{\mu=1}^4(D_\mu-x_\mu)^2+\frac{\Vh}{2\pi}\sum_{m=1}^N|\vec\rho_m|\,\delta^{(2)}(\xi-\mu_m).
    \label{eq:app-K0}
\end{align}
The three self-duality conditions in \eqref{eq:adhm-sd}, $\tr_2(\sigma_l\Delta^\dagger\Delta)=0$, are $\mathcal{K}^{l}_x=0$, which is precisely the Nahm/Hitchin equation with impurities \eqref{eq:nahm-general}. Note that $\mathcal{K}^{l}_x$ is independent of $x$, as it should be.
For $Q_{\mathrm{top}}=1$, the $\Ah_\mu(\xi)$ are functions and \eqref{eq:nahm-general} reduces to \eqref{eq:nahm}: the $l=3$ component reads $\Fh_{43}=-\Fh_{34}=\frac{\Vh}{2\pi}\sum_m\rho^3_m\delta^{(2)}(\xi-\mu_m)$ since $\Fh_{12}=0$, and the left-hand sides of the $l=1,2$ components, $\partial_{\xi_4}\Ah_1-\partial_{\xi_3}\Ah_2$ and $\partial_{\xi_3}\Ah_1+\partial_{\xi_4}\Ah_2$, combine into $\frac12(\partial_{\xi_3}-\im\partial_{\xi_4})\hat H$ with $\hat H=\Ah_1+\im\Ah_2$.

Once \eqref{eq:nahm-general} holds, $\mathcal{K}_x=\sigma_0\,\mathcal{K}^{0}_x$, and \eqref{eq:adhm-sd} identifies $\mathcal{K}^{0}_x$ with the transform of $f_x^{-1}$. The relation $f_x^{-1}f_x=\mathbbm{1}_k$ then becomes, by the product rule, $\mathcal{K}^{0}_x\hat f_x(\xi,\xi')=\Vh\,\delta^{(2)}(\xi-\xi')$, which is \eqref{eq:fhat}. Finally, inverting \eqref{eq:FT}, $f_{IJ}=\Vh^{-2}\int\diff^2\xi\,\diff^2\xi'\,\ee^{-\im(I\cdot(L\xi)-J\cdot(L\xi'))}\hat f_x(\xi,\xi')$, together with $\lambda_I=\sum_m\ee^{\im I\cdot(L\mu_m)}\Pi_m\zeta$, gives
\begin{align}
    \lambda\,[\sigma_\mu\otimes f_x]\,\lambda^\dagger
    &=\sum_{I,J}\lambda_I\,\sigma_\mu\,(f_x)_{IJ}\,\lambda_J^\dagger
    =\sum_{m,m'=1}^N\hat f_x(\mu_m,\mu_{m'})\,\Pi_m\zeta\,\sigma_\mu\,\zeta^\dagger\Pi_{m'} ,
\end{align}
whose $\mu=4$ and $\mu=j$ components, combined with \eqref{eq:def_varphi}, yield \eqref{eq:phi_f} and \eqref{eq:eta_f}, respectively.

\bibliographystyle{utphys}
\bibliography{./QFT,./refs}

\end{document}